\documentclass[journal]{IEEEtran}

\usepackage{graphicx}
\usepackage{cite}
\usepackage{float}
\usepackage{amsmath}
\usepackage{amssymb}
\usepackage{pgfplots}
\pgfplotsset{compat=newest}
\usepackage[caption=false,font=footnotesize]{subfig}
\usepackage[hidelinks]{hyperref}
\usepackage{tikz}
\usepackage{tabularx}
\usepackage{booktabs}
\usepackage{hyphenat}
\usepackage{siunitx}
\usepackage{enumitem}
\usepackage{bbm}
\usepackage{dblfloatfix}
\usepackage{balance}
\usepackage{xspace}
\usetikzlibrary{arrows.meta, positioning}
\DeclareSIUnit{\molar}{M}
\newcommand{\DA}{\ensuremath{\mathrm{DA}}\xspace}
\newcommand{\HT}{\ensuremath{\mathrm{5\text{-}HT}}\xspace}
\newcommand{\CTRL}{\ensuremath{\mathrm{CTRL}}\xspace}

\makeatletter
\def\@IEEEBIOskipN{1\baselineskip}
\renewenvironment{IEEEbiography}[2][]{\normalfont\@IEEEcompsoconly{\sffamily}\footnotesize%
\unitlength 1in\parskip=0pt\par\parindent 1em\interlinepenalty500%
\@IEEEtrantmpdimenA=\@IEEEBIOhangdepth%
\advance\@IEEEtrantmpdimenA by \@IEEEBIOskipN%
\advance\@IEEEtrantmpdimenA by 1\baselineskip%
\@IEEEtranneedspace{\@IEEEtrantmpdimenA}{\relax}%
\vskip \@IEEEBIOskipN%
\def\@IEEEtempbiographybox{{\setlength{\fboxsep}{0pt}\framebox{%
\begin{minipage}[b][\@IEEEBIOphotodepth][c]{\@IEEEBIOphotowidth}\centering PLACE\\ PHOTO\\ HERE \end{minipage}}}}%
\@ifmtarg{#1}{\relax}{\def\@IEEEtempbiographybox{\mbox{\begin{minipage}[b][\@IEEEBIOphotodepth][c]{\@IEEEBIOphotowidth}%
\centering%
#1%
\end{minipage}}}}%
\if@IEEEbiographyTOCentrynotmade%
\setcounter{IEEEbiography}{-1}%
\refstepcounter{IEEEbiography}%
\addcontentsline{toc}{section}{Biographies}%
\global\@IEEEbiographyTOCentrynotmadefalse%
\fi%
\refstepcounter{IEEEbiography}%
\addcontentsline{toc}{subsection}{#2}%
\let\@IEEEBIOORGparCMD=\par%
\edef\par{\hfil\break\indent}%
\settoheight{\@IEEEtrantmpdimenA}{\@IEEEtempbiographybox}%
\@IEEEtrantmpdimenB=\@IEEEBIOhangdepth%
\@IEEEtrantmpcountA=\@IEEEtrantmpdimenB%
\divide\@IEEEtrantmpcountA by \baselineskip%
\advance\@IEEEtrantmpcountA by 1%
\hangindent\@IEEEBIOhangwidth%
\hangafter-\@IEEEtrantmpcountA%
\settoheight{\@IEEEtrantmpdimenB}{\mbox{T}}%
\noindent\makebox[0pt][l]{\hspace{-\@IEEEBIOhangwidth}\raisebox{\@IEEEtrantmpdimenB}[0pt][0pt]{%
\raisebox{-\@IEEEBIOphotodepth}[0pt][0pt]{\@IEEEtempbiographybox}}}%
\noindent\textbf{#2\ }\@IEEEgobbleleadPARNLSP}{\relax\let\par=\@IEEEBIOORGparCMD\par%
\ifnum \prevgraf <\@IEEEtrantmpcountA\relax%
    \advance\@IEEEtrantmpcountA by -\prevgraf%
    \advance\@IEEEtrantmpcountA by -1\relax%
    \@IEEEtrantmpdimenA=\baselineskip%
    \multiply\@IEEEtrantmpdimenA by \@IEEEtrantmpcountA%
    \noindent\rule{0pt}{\@IEEEtrantmpdimenA}%
\fi%
\par\normalfont}
\makeatother

\title{A Control-Referenced Tri-Channel OECT Receiver for Hybrid Molecular Communication Toward Brain Organoid Interfaces}
\author{Hongbin Ni, \textit{Student Member, IEEE} and Ozgur B. Akan, \textit{Fellow, IEEE}%
\thanks{The authors are with the Internet of Everything Group, Electrical Engineering Division, Department of Engineering, University of Cambridge, CB3 0FA Cambridge, U.K. (e-mail: hn345@cam.ac.uk).}%
\thanks{Ozgur B. Akan is also with the Center for neXt-Generation Communications (CXC), Department of Electrical and Electronics Engineering, Koc University, 34450 Istanbul, Turkey (e-mail: oba21@cam.ac.uk, akan@ku.edu.tr).}%
\thanks{This work was supported in part by the AXA Research Fund (AXA Chair for Internet of Everything at Ko\c{c} University).}
}

\begin{document}

\maketitle

\begin{abstract}
Brain organoid interfaces that seek neuromodulator readout benefit from chemical receivers with molecular specificity and tolerance to drift. This paper presents a receiver-centric theoretical study of a control-referenced tri-channel organic electrochemical transistor (OECT) receiver with dopamine and serotonin selective pixels alongside a hydrogel-matched control pixel. The Ag/AgCl electrode provides the electrochemical gate reference, whereas the control pixel is used only as a matched reference for common-mode drift and other low-frequency baseline fluctuations during amplitude decisions. We couple finite-duration release, restricted diffusion with clearance, aptamer binding, OECT transduction, and correlated thermal, flicker, and drift noise, and we evaluate MoSK, CSK-4, and a 2-bit Hybrid detector on the same front-end by Monte Carlo simulation. At $r=\SI{45}{\micro\meter}$, control referencing mainly benefits the Hybrid amplitude branch, reducing Hybrid SER from $3.71\times 10^{-2}$ to $1.09\times 10^{-2}$ at $N_m=1.40\times 10^4$ molecules/symbol while barely changing the MoSK component. In calibrated no-ISI front-end benchmarks, Hybrid+CTRL reaches an LoD of 11866 molecules/symbol at \SI{45}{\micro\meter} and remains below CSK-4+CTRL over much of the medium-to-long-distance range studied. The reported SER and LoD values are scenario-based receiver forecasts, whereas the more transferable result is the regime-dependent rule for when matched control referencing benefits Hybrid amplitude decoding.
\end{abstract}


\section{Introduction}
\label{sec:introduction}

\IEEEPARstart{B}{rain} organoids are self-organized three-dimensional neural tissues derived from pluripotent stem cells~\cite{lancaster2013cerebral}. They exhibit emergent activity patterns, including synchronous bursting and oscillations, and support mechanistic studies and closed-loop interfacing across weeks to months of maturation~\cite{trujillo2019complex}. Beyond electrophysiology, organoid state is strongly influenced by chemical signaling. Neuromodulators regulate network excitability and plasticity. Dopamine (DA) in particular shapes reinforcement learning and gain control over multiple timescales~\cite{schultz2007multiple}. Serotonin (5-HT) is another major modulatory transmitter with pronounced release and uptake dynamics~\cite{bunin1998release}. Organoid-facing interfaces therefore benefit from sensing neuromodulators with molecular specificity.

Electrochemical monitoring of neuromodulators faces well-known tradeoffs. Fast-scan cyclic voltammetry offers millisecond-scale temporal resolution but relies on large voltage excursions and background subtraction and can struggle with selectivity in complex mixtures~\cite{heien2004real}. Microdialysis provides chemical specificity through offline analysis but is invasive and slow, and it perturbs the extracellular milieu~\cite{chefer2009overview}. Optical reporters require illumination and can be limited by photobleaching and phototoxicity, especially in long-duration live imaging experiments~\cite{combs2010fluorescence}. These constraints motivate low-bias, label-free receiver architectures that are compatible with long-term organoid experiments and that can be analyzed within a molecular communication (MC) framework~\cite{akyildiz2015internet,farsad2016comprehensive}.

Organic electrochemical transistors (OECTs) and aptamers provide a promising receiver substrate. OECTs transduce ionic charge in an electrolyte into electronic current through volumetric doping of mixed ionic-electronic conductors such as PEDOT:PSS, enabling high transconductance at sub-\SI{600}{mV} biases~\cite{rivnay2018oect,khodagholy2013high,friedlein2018physics}. OECTs have been demonstrated as on-site amplifiers for electrochemical aptamer-based sensing, improving sensitivity for miniaturized electrodes~\cite{ji2023eab_oect}, and dopamine OECT aptasensors have also been reported~\cite{soliman2024dopamineoect}. Experimental multiplexed dopamine/serotonin aptamer transistor probes further support the broader feasibility of simultaneous DA/5-HT transistor sensing in complex media, although not with a matched control axis or an OECT front-end~\cite{gao2022multiplexed}. Aptamers can recognize small molecules under physiological ionic strength and can yield species-dependent electrostatic transduction through binding-induced conformational changes~\cite{nakatsuka2018aptamer}. Notably, dopamine and serotonin aptamer FETs have been reported to exhibit opposite-signed source--drain current responses (dopamine decreases $I_{DS}$ whereas serotonin increases $I_{DS}$), attributed to opposite binding-induced reorientation of the negatively charged aptamer backbone toward or away from the semiconductor channel~\cite{nakatsuka2018aptamer}. We therefore treat the sign difference between DA and 5-HT coupling as a receiver-level modeling assumption motivated by aptamer-FET evidence, rather than as an experimentally verified property of the exact tri-channel OECT stack considered here.

For MC receiver design, two challenges are central at organoid-relevant length scales. Firstly, signal formation depends on the combined effects of diffusion in restricted extracellular space, clearance, and binding kinetics. Extracellular space parameters such as volume fraction $\alpha\approx 0.2$ and tortuosity $\lambda\approx 1.6$ substantially reshape impulse responses and should be included to avoid optimistic transport predictions~\cite{sykova2008diffusion,nicholson1998diffusion}. Secondly, OECT measurements are often limited by low-frequency drift and correlated $1/f$ noise that are shared across devices in a common electrolyte and bias environment~\cite{stoop2017noise,friedlein2018physics}. Differential referencing, matched to the same electrolyte and bias conditions, is widely used to suppress common-mode drift in biosensing readout chains~\cite{hanzaee2025differential,uguz2024complementary,song2024drift}, but it has not been systematically integrated into a multi-species MC receiver model with matched decision statistics.

Prior receiver-centric MC studies have largely focused on single-axis biosensor receivers or on cross-reactive receptor arrays without a matched control axis~\cite{kuscu2016physical,kuscu2016sinw,kuscu2021graphene,civas2024freqdetect}. Recent organoid-focused receiver analysis has also considered a 3D biosensor receiver for Internet-of-Organoids settings~\cite{zhang2025organoid3d}, and recent control-theoretic work has explored symbol-by-symbol adaptive thresholding in a BioFET-oriented MC receiver (ART-Rx)~\cite{ni2025artrx}. Hybrid modulation itself is not new in MC, with prior work on concentration-time hybrid modulation~\cite{gursoy2021conctime} and MoSK-CSK combinations in multi-link settings~\cite{sabu2019hybridmoskcsk}. The missing piece here is different: not adaptive thresholding on a single decision axis, but a control-referenced tri-channel OECT front-end that makes an identity-plus-amplitude Hybrid symbol analyzable under organoid-like restricted transport, common-mode drift, and low-frequency fluctuations. This paper addresses that gap with a receiver-centric approach.

\begin{itemize}
  \item \textit{Tri-channel OECT receiver architecture}: We propose a tri-channel OECT receiver for organoid wells with two selective aptamer-gated channels (DA and 5-HT) and a hydrogel-matched control channel (CTRL). The Ag/AgCl electrode fixes the electrochemical gate reference. The control channel is not treated as a third analyte axis. Instead, it serves as a matched nuisance reference for amplitude decisions under shared drift and correlated low-frequency noise.
  \item \textit{Unified receiver model}: We develop a physics-based model that maps finite-duration molecular release through restricted diffusion and clearance to Langmuir binding, optionally with a Damk\"ohler-inspired effective on-rate, and then to OECT transduction under an explicit correlated thermal, flicker, and drift noise model in which the low-frequency components are partially shared across channels while thermal noise is treated as uncorrelated.
  \item \textit{Hybrid-centric evaluation on one front-end}: We treat the 2-bit Hybrid scheme as the primary target and use MoSK and CSK-4 as informative limiting cases that expose the molecule-identity and amplitude branches of the same receiver. This lets us attribute gains to the tri-channel front-end rather than to unrelated platform changes.
\end{itemize}

Accordingly, the SER and LoD values reported below are best interpreted as scenario-based receiver forecasts under the baseline assumptions, whereas the more transferable result is the receiver-design rule for when matched control referencing benefits the Hybrid amplitude branch.

All numerical figures in Sections~\ref{sec:results} and~\ref{sec:discussion} are direct outputs of custom Monte Carlo simulations implementing the receiver model in Sections~\ref{sec:system} and~\ref{sec:analytical}, with threshold calibration and confidence-controlled seed schedules detailed in Section~\ref{sec:simulation}. The remainder of the paper is organized as follows. Section~\ref{sec:system} describes the system architecture, transport setting, and modulation formats. Section~\ref{sec:analytical} presents the analytical model and decision statistics. Section~\ref{sec:simulation} details the simulation framework and parameterization used for numerical evaluation. Section~\ref{sec:results} reports the numerical results. Section~\ref{sec:discussion} discusses the main receiver-design insights, relation to prior work, and limitations. Section~\ref{sec:conclusion} concludes the paper.

\section{System and Architecture}
\label{sec:system}

\subsection{Tri-Channel OECT Receiver and Physical Layout}

\begin{figure*}[!t]
  \centering
  \subfloat[Three-dimensional conceptual view of the organoid well and the tri-channel OECT receiver.]{
    \includegraphics[width=0.48\textwidth]{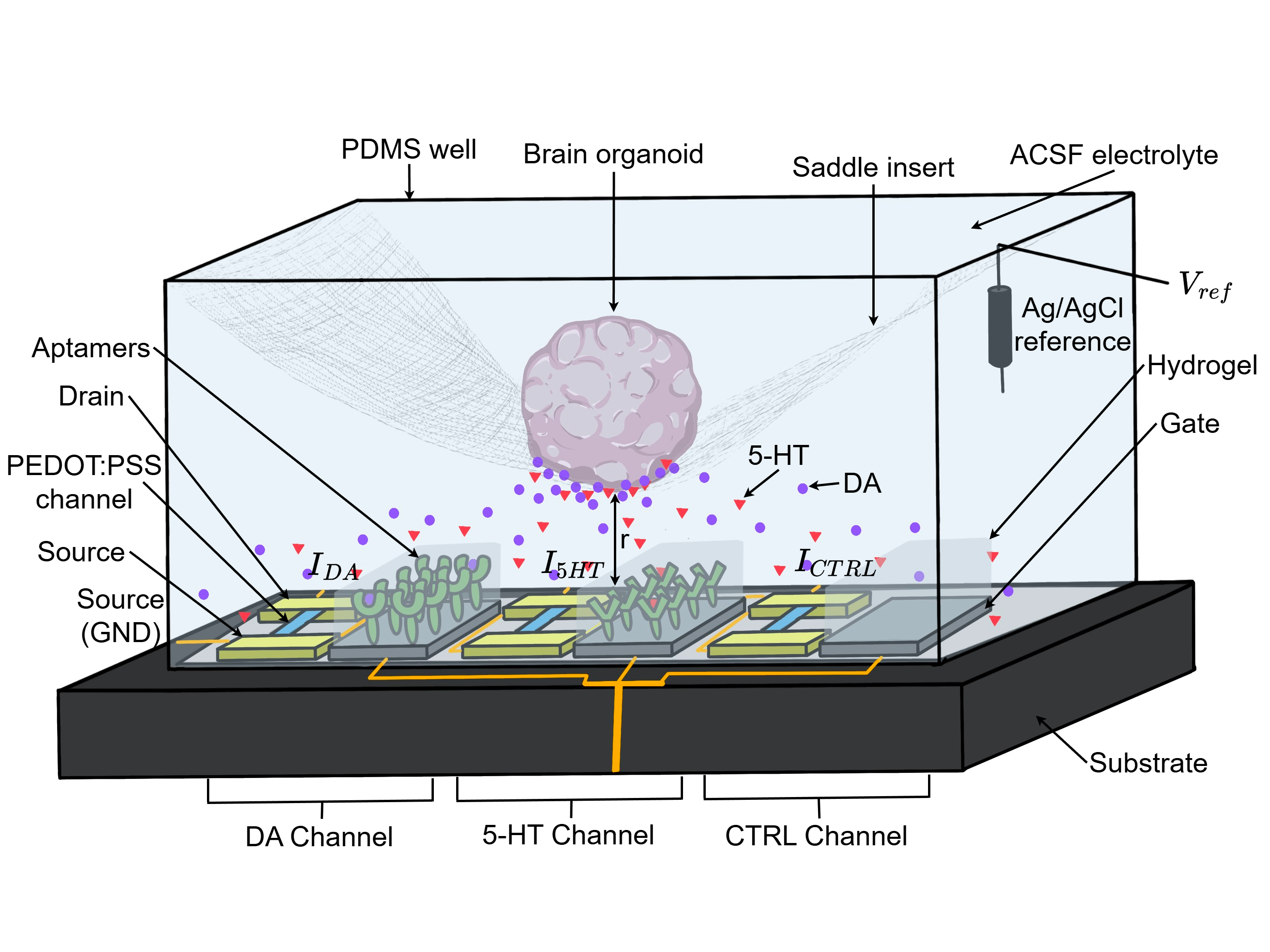}
    \label{fig:device3d}
  }
  \hfill
  \subfloat[Two-dimensional cross-section showing the organoid-to-gate gap $r$, the hydrogel layers, and the three OECT pixels (DA, 5-HT, CTRL) sharing a common gate bias referenced to Ag/AgCl.]{
    \includegraphics[width=0.48\textwidth]{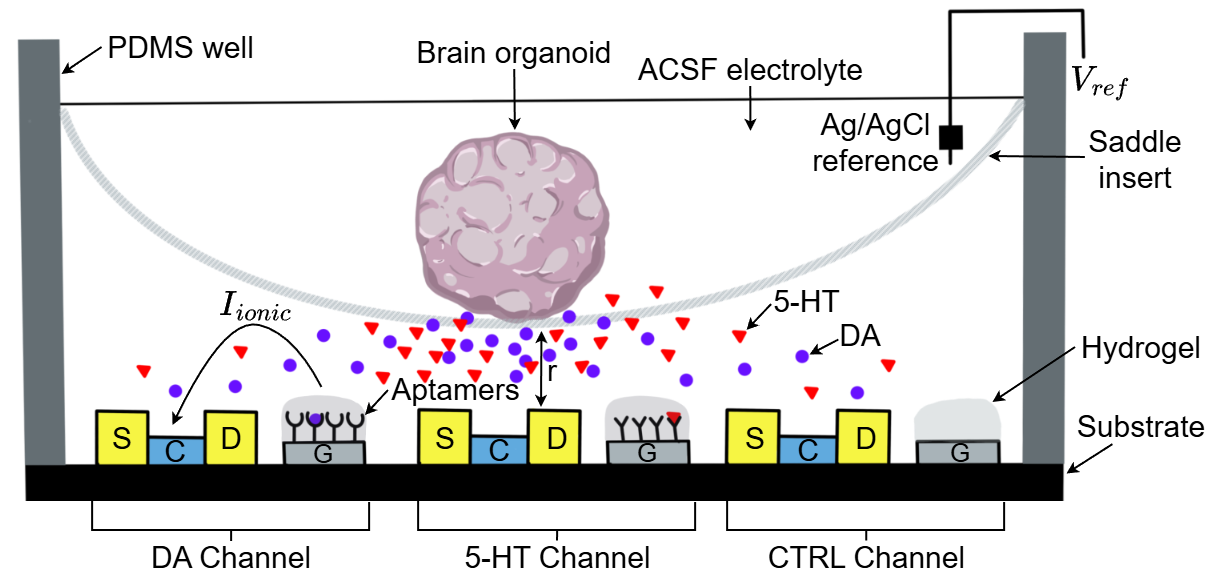}
    \label{fig:device2d}
  }
  \caption{Tri-channel OECT receiver concept for brain organoid molecular communication. (a) Three-dimensional conceptual view of the organoid well and tri-channel OECT array. (b) Cross-section highlighting the organoid-to-gate separation $r$, hydrogel and aptamer functionalization, and shared gate bias referenced to Ag/AgCl. Two selective channels target \DA and \HT, while a hydrogel-matched control channel (\CTRL) provides common-mode referencing for drift and correlated low-frequency noise.}
  \label{fig:device_overview}
\end{figure*}

Figure~\ref{fig:device_overview} illustrates the proposed receiver. Three square gold micro-gate pads of side length \SI{200}{\micro\meter} are patterned on a common substrate in a planar tri-channel layout. Each gate pad is associated with a PEDOT:PSS OECT channel and an independent drain readout, forming three sensing pixels corresponding to \DA, \HT, and \CTRL. The gate electrodes share a common bias node $V_g$ referenced to an Ag/AgCl electrode in the bath. This shared electrolyte and bias environment introduces common-mode drift and low-frequency fluctuations across channels. Differential and complementary OECT front-ends exploit this common-mode structure for rejection using subtraction or differential readout.~\cite{hanzaee2025differential,uguz2024complementary,song2024drift}

Each gate pad is covered by a porous hydrogel layer (thickness \SIrange{5}{20}{\micro\meter}) that protects the functional layer while allowing rapid ionic transport and maintaining high water content (typically 70--99\% water for hydrogels)~\cite{li2016hydrogel}. The \DA and \HT gates are functionalized with nucleic-acid aptamers selective to the corresponding neuromodulator~\cite{nakatsuka2018aptamer}. The \CTRL gate carries the same hydrogel but no aptamer. It is therefore not a third analyte sensor. Instead, it measures the same electrolyte, bias, and hydrogel background seen by the selective channels without specific binding and serves as a matched reference for amplitude decisions~\cite{song2024drift,hanzaee2025differential}. Throughout this paper, control referenced means subtraction against this matched \CTRL pixel, not against the Ag/AgCl electrode.

A brain organoid is suspended in artificial cerebrospinal fluid (ACSF) inside a PDMS well and positioned above the three gates. A saddle insert or sparse inert micro-spacers constrain the separation between the organoid surface and the gate plane to a controlled gap $r$ in the range \SIrange{25}{130}{\micro\meter} (implemented via a saddle insert or spacers). In the numerical evaluation in Section~\ref{sec:simulation} we sweep $r\in\{25,30,35,40,45,60,75,90,110,130\}\,\si{\micro\meter}$ to span plausible organoid-to-gate spacings. For modeling and evaluation, we treat organoid release as an effective point source located above the receiver triad and, unless otherwise stated, we use a common separation $r$ for all three channels to isolate the receiver front-end and detection effects under a controlled transport condition. The same model supports channel-specific separations $r_{\DA}$, $r_{\HT}$, and $r_{\CTRL}$ to capture lateral offsets or heterogeneous placement, but these are not required for the receiver-centric comparisons in this paper. This common-distance geometry is therefore a deliberate receiver-centric idealization rather than the default physical placement expected in an organoid well.

\begin{table}[t]
  \centering
  \caption{Nominal geometry, materials, and operating points for the tri-channel OECT receiver}
  \label{tab:device_geom}
  \begin{tabularx}{\columnwidth}{@{}p{0.43\columnwidth}X@{}}
    \toprule
    Item & Nominal value or choice \\
    \midrule
    PEDOT:PSS channel & length \SIrange{60}{100}{\micro\meter}, width \SIrange{10}{20}{\micro\meter}, thickness \SIrange{100}{500}{nm}~\cite{khodagholy2013high,rivnay2015thickness,friedlein2018physics} \\
    Gate pads (Ti/Au) & area $200\times 200~\si{\micro\meter\squared}$, pitch \SIrange{250}{400}{\micro\meter} (DA, 5-HT, CTRL) \\
    Hydrogel over gates & porous PEG--DA or alginate, thickness \SIrange{5}{20}{\micro\meter}, water content typically 70--99\%~\cite{li2016hydrogel} \\
    Organoid spacing & assumed gap $r$ between \SIrange{25}{130}{\micro\meter} (set by spacer/gel thickness), swept to span plausible chip geometries and extracellular diffusion length scales~\cite{sykova2008diffusion,nicholson1998diffusion} \\
    PDMS well & inner diameter \SIrange{3}{6}{mm}, height \SIrange{1}{3}{mm}, ACSF volume \SIrange{200}{500}{\micro\liter} \\
    OECT operating point & $g_m \approx \SI{5}{mS}$, $C_{\mathrm{tot}}\approx \SI{50}{nF}$, $I_{\mathrm{DC}}\approx \SI{100}{\micro A}$, $V_g \approx \SI{-0.2}{V}$~\cite{khodagholy2013high,rivnay2015thickness,rivnay2018oect,hanzaee2025differential} \\
    \CTRL channel & hydrogel only, no aptamer, matched geometry to DA and 5-HT \\
    \bottomrule
  \end{tabularx}
\end{table}

\subsection{Electronic Readout and Receiver Processing}

\begin{figure*}[!t]
  \centering
  \subfloat[Analog front-end and readout.]{
    \includegraphics[width=0.88\textwidth]{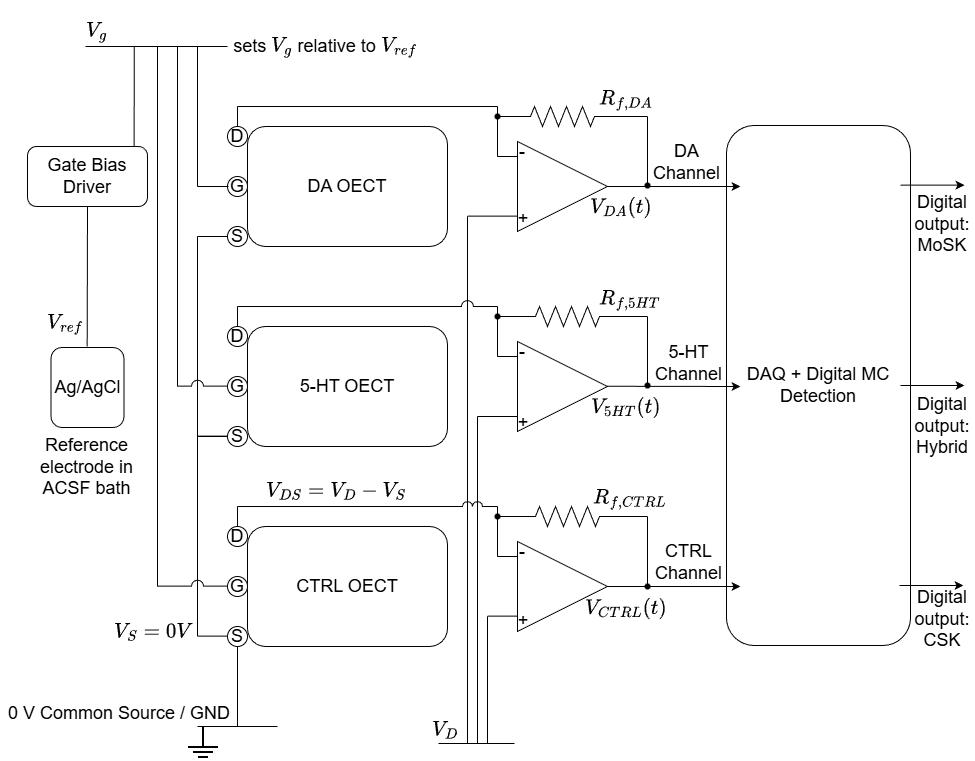}
    \label{fig:signalchain_full}
  }\\[1mm]
  \subfloat[Companion digital decision logic.]{
    \includegraphics[width=0.88\textwidth]{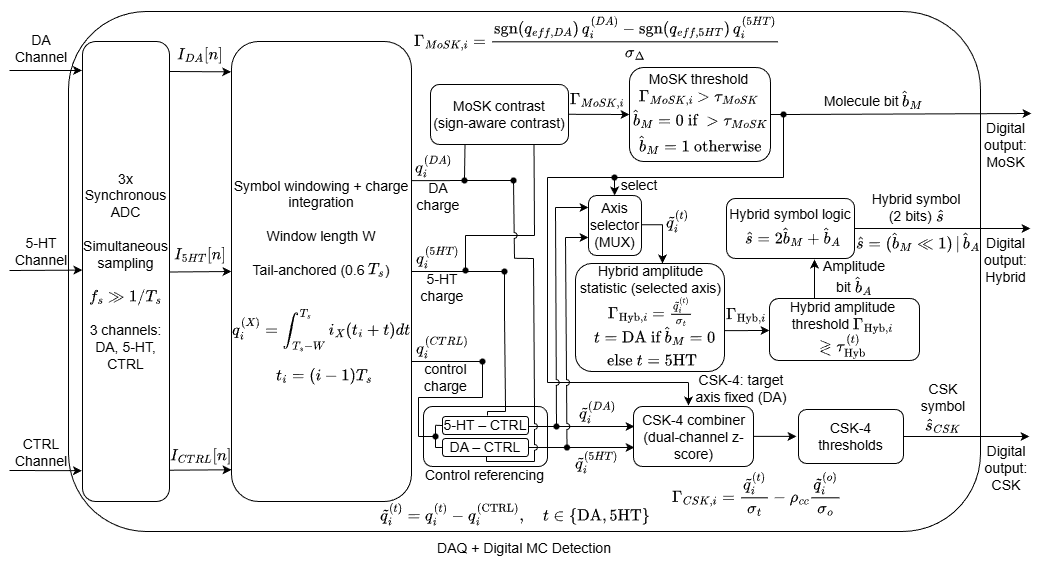}
    \label{fig:signalchain_zoom}
  }
  \caption{Tri-channel acquisition and detection. (a) Analog front-end and readout. (b) Companion digital processing and decision statistics. Synchronous sampling produces three drain-current traces per symbol. Tail-anchored windowing converts currents to charge statistics. MoSK uses a sign-aware contrast between \DA and \HT without control subtraction. CSK-4 applies control referencing and a dual-channel variance-normalized combiner before thresholding. The hybrid detector first decides molecule identity using MoSK and then applies a single-axis control-referenced amplitude threshold on the selected species axis. The DAQ/digital block is algorithmic and can be implemented off chip or in a mixed-signal backend. It is not intended to imply monolithic on-array digital integration.}
  \label{fig:signal_chain}
\end{figure*}

Figure~\ref{fig:signal_chain} should be read as a receiver abstraction. The OECT receiver hardware comprises the tri-channel analog sensing front-end and readout path up to the acquisition of the drain-current traces. The blocks in Fig.~\ref{fig:signal_chain}(b) denote the companion digital processing used to compute the decision statistics in Section~\ref{sec:analytical}. They can be implemented off chip in a DAQ, microcontroller, or mixed-signal backend and are not intended to imply large-scale digital logic monolithically integrated inside the OECT pixel array. A common gate bias $V_g$ is applied relative to an Ag/AgCl reference electrode, while the three OECT drain currents are read out independently under a shared source potential. Over symbol interval $i$ of length $T_s$, the receiver forms charge statistics by tail-anchored integration over a window of length $W=\eta T_s$. In this paper, control referenced explicitly means subtracting the simultaneously measured \CTRL current, or equivalently the charge integrated over the same decision window, from a selective-channel measurement before an amplitude decision is made. MoSK does not use control subtraction. CSK-4 applies control referencing and a dual-channel variance-normalized combiner, while the hybrid detector first resolves molecule identity using MoSK and then applies a single-axis control-referenced amplitude threshold on the selected species axis.

\subsection{Neuromodulator Source and Transport Path}

For communication modeling, the organoid is treated as a source that releases a finite burst of \DA or \HT molecules when stimulated. Monoamine release and uptake kinetics can be heterogeneous across microdomains and time-varying across repeated stimuli, motivating a scenario-based approach\cite{lippert2019time,taylor2015kinetic}. Molecules propagate through restricted extracellular space, modeled through an effective volume fraction $\alpha$ and tortuosity $\lambda$\cite{sykova2008diffusion}. We include an effective first-order sink $k_{\mathrm{clear}}$ capturing net removal from the accessible extracellular space (diffusive loss from the modeled volume plus uptake and metabolism when present). This first-order sink can be interpreted as a local linearization of Michaelis--Menten uptake in the sub-saturation regime, $k_{\mathrm{clear}}\approx V_{\max}/K_m$, which is commonly used in dopamine kinetic models\cite{walters2014restricted,wightman1988dopamine}. Using commonly used striatal dopamine uptake parameters from Fast-scan cyclic voltammetry (FSCV)-based modeling ($V_{\max}\approx 5~\mu\mathrm{M}/\mathrm{s}$, $K_m\approx 0.2~\mu\mathrm{M}$) yields a pseudo-first-order rate $V_{\max}/K_m\approx 25~\mathrm{s}^{-1}$\cite{wightman1988dopamine}. Conversely, uptake inhibition (e.g., dopamine transporter blockade) can yield substantially slower, diffusion-dominated decay tails\cite{hoffman2016transport}. Because organoid transporter expression and geometry can vary widely, we treat $k_{\mathrm{clear}}$ as a tunable scenario parameter and, for the baseline organoid setting, adopt a slow-clearance regime $k_{\mathrm{clear}}=\SI{0.01}{s^{-1}}$ to conservatively model long extracellular tails (worst-case ISI). Although the gate hydrogel films add a thin polymer network at the interface, hydrogels are predominantly water (typically 70--99\% water) and small molecules diffuse with limited hindrance when the hydrogel mesh size is large relative to the solute\cite{li2016hydrogel}. We therefore model the transport path as a single effective medium, absorbing residual deviations from bulk diffusion into the effective $\lambda$ and $k_{\mathrm{clear}}$.

\subsection{Modulation Formats and Energy Normalization}

We consider three modulation formats over the same tri-channel receiver. The Hybrid 2-bit scheme is the main target. MoSK and CSK-4 are retained as supportive baselines that expose the identity-only and amplitude-only limits of the same front-end.

\begin{itemize}
  \item \textit{MoSK}: each symbol uses either \DA or \HT, encoding one bit in molecule identity. Opposite-signed effective gating charges on the two aptamer channels yield opposite-signed drain-current responses, enabling sign-aware molecule discrimination.
  \item \textit{CSK-4}: a single molecule axis carries information through four amplitude levels. We use \DA as the default CSK axis, while the framework applies identically to \HT. Three calibrated thresholds partition the control-referenced amplitude statistic into four classes, yielding two bits per symbol.
  \item \textit{Hybrid (2-bit)}: the most significant bit selects \DA versus \HT, while the least significant bit selects low versus high amplitude on the selected species axis. Detection first resolves molecule identity using the MoSK statistic and then resolves the amplitude bit on the selected axis.
\end{itemize}

For fair comparisons across modulation formats, we keep the average molecule budget per symbol fixed at $N_m$ and scale peak amplitudes accordingly. For CSK-4, symbol levels map to fractions $\{1/4,2/4,3/4,1\}$ of a peak value $N_m^{\mathrm{pk}}$ such that the mean across equally likely symbols equals $N_m$. For the hybrid scheme, the low and high amplitude levels are $\{0.5,1\}$ of $N_m^{\mathrm{pk}}$ with the same mean constraint. Molecular shot noise is modeled by drawing the realized emitted molecule count from a Poisson distribution with mean equal to the selected level~\cite{pierobon2011diffusion}.

\subsection{Symbol Timing, Decision Window, and ISI Control}

Each symbol occupies a period $T_s$. The receiver integrates the drain currents over a decision window of length $W \le T_s$ to form charge statistics. Throughout this work, the decision window is a tail-anchored fraction of the symbol interval,
\begin{equation}
  W = \eta T_s,\quad \eta = 0.6,
  \label{eq:window_fraction}
\end{equation}
which captures late-time signal energy while reducing sensitivity to early transients~\cite{farsad2016comprehensive}.

Inter-symbol interference (ISI) arises from residual molecules in the medium and from slow unbinding of occupied aptamers. We model ISI by superposing residual concentration contributions from a finite history of past symbols. The guard factor $g$ controls the tradeoff between symbol rate and ISI by increasing the symbol period relative to a diffusion-based characteristic timescale. For numerical evaluation, we write
\begin{align}
  t_{\mathrm{char}}(r) &\triangleq c_t\,\frac{r^2}{D_{\mathrm{eff}}}, \\
  T_s(r) &= \max\!\left(T_{\min},\,\left\lceil \frac{(1+g)t_{\mathrm{char}}(r)}{\Delta t}\right\rceil \Delta t\right),
  \label{eq:dynamic_ts}
\end{align}
where $\Delta t$ is the simulation time step and $T_{\min}$ is a minimum symbol period. Here $D_{\mathrm{eff}}=D/\lambda^2$~\cite{sykova2008diffusion}, and $c_t>0$ is a conservative design constant rather than a canonical transport identity. In the baseline numerical evaluation we set $c_t=3$, chosen to produce a deliberately conservative diffusion timescale for guard-factor design. For the no-clearance Green's function in \eqref{eq:greens}, the impulse-response peak would instead occur at $r^2/(6D_{\mathrm{eff}})$. We use the slower of the two neuromodulator diffusivities (\DA in Table~\ref{tab:sim-params}) for timing so that $T_s(r)$ is conservative for both \DA and \HT. The ISI memory depth is selected automatically from the window coefficients in Section~\ref{sec:analytical} so that the residual diffusion tail contribution is negligible relative to the within-symbol response, and it is capped to limit computation in long-tail regimes. To reflect slow aptamer unbinding in guard-factor sweeps, we optionally use a binding-aware symbol-period policy in which the diffusion-based timescale $t_{\mathrm{char}}(r)$ in \eqref{eq:dynamic_ts} is replaced by $\max\{t_{\mathrm{char}}(r),\,\kappa/\bar{k}_{\mathrm{off}}\}$, where $\kappa=5$ (so that $\exp(-\kappa)\approx0.7\%$, i.e., residual binding after $\kappa$ time constants is negligible) and $\bar{k}_{\mathrm{off}}$ is the average off-rate across \DA and \HT.

\section{Analytical Modeling}
\label{sec:analytical}

\subsection{End-to-End Signal Path and Charge Statistics}

During the $i$th symbol interval, the drain current on channel $X\in\{\DA,\HT,\CTRL\}$ is denoted $i_X(t)$ for $t\in[(i-1)T_s,\,iT_s)$. The receiver maps each current trace to a scalar charge statistic by integrating over a decision window of length $W\le T_s$ that is anchored at the end of the symbol interval,
\begin{equation}
  q_i^{(X)} = \int_{T_s-W}^{T_s} i_X(t_i+t)\,\mathrm{d}t,\quad t_i=(i-1)T_s.
  \label{eq:charge_stat}
\end{equation}
For amplitude decisions, control referencing is applied at the waveform level by subtracting the \CTRL current before window integration, while the MoSK contrast in \eqref{eq:gamma_mosk} is formed directly from the two selective channels. The objective of the model is to relate $\{q_i^{(X)}\}$ to the transmitted symbol sequence while accounting for restricted diffusion, binding kinetics, OECT transduction, and correlated electrical noise.
\subsection{Decision Statistics for MoSK, CSK, and Hybrid}

\paragraph{MoSK statistic}
MoSK decides between \DA and \HT using a sign-aware differential charge contrast. Unless otherwise stated, we operate in a variance-normalized (z-score) domain so that thresholds are dimensionless and comparable across operating points. The MoSK decision statistic is
\begin{equation}
  \Gamma_{\mathrm{MoSK},i}
  = \frac{\mathrm{sgn}(q_{\mathrm{eff},\DA})\,q_i^{(\DA)}
    - \mathrm{sgn}(q_{\mathrm{eff},\HT})\,q_i^{(\HT)}}{\sigma_{\Delta}},
  \label{eq:gamma_mosk}
\end{equation}
where $\sigma_{\Delta}$ is the standard deviation of the raw sign-aware difference under the channel noise model. With opposite-signed effective charges, $\Gamma_{\mathrm{MoSK},i}$ is positive under \DA transmission and negative under \HT transmission in the noise-free limit. The receiver compares $\Gamma_{\mathrm{MoSK},i}$ to a calibrated threshold $\tau_{\mathrm{MoSK}}$. Under symmetric noise and equal priors, $\tau_{\mathrm{MoSK}}=0$ is the maximum-likelihood threshold. Although opposite-signed coupling motivates this particularly simple statistic, the receiver concept does not require strict sign inversion. If a future OECT implementation yields same-sign but still selective \DA and \HT responses, molecule identity can instead be resolved from the bivariate vector $[q_i^{(\DA)},q_i^{(\HT)}]^\top$ using an equal-covariance Gaussian discriminant between the two molecule classes, while the control-referenced amplitude step in \eqref{eq:hyb_amp_stat} remains unchanged.

\paragraph{Control-referenced charges}
For amplitude decisions, the receiver forms control-referenced charges on the selective channels by subtracting the hydrogel-matched \CTRL channel,
\begin{equation}
  \tilde{q}_i^{(t)}
  = q_i^{(t)} - q_i^{(\CTRL)},\quad t\in\{\DA,\HT\}.
  \label{eq:qdiff}
\end{equation}
In implementation, subtraction is applied at the waveform level prior to window integration, which is equivalent to \eqref{eq:qdiff} when the same decision window is used across channels.

\paragraph{Dual-channel CSK statistic}
Let $\tilde{q}_i^{(t)}$ be the control-referenced charge on the target axis and let $\tilde{q}_i^{(o)}$ be the corresponding control-referenced charge on the other selective axis. With standard deviations $\sigma_t,\sigma_o$ and residual correlation coefficient $\rho_{\mathrm{cc}}$ between $\tilde{q}_i^{(t)}$ and $\tilde{q}_i^{(o)}$, we use the variance-normalized contrast
\begin{equation}
  \Gamma_{\mathrm{CSK},i}
  = \frac{\tilde{q}_i^{(t)}}{\sigma_t}
    - \rho_{\mathrm{cc}}\frac{\tilde{q}_i^{(o)}}{\sigma_o}.
  \label{eq:csk_stat}
\end{equation}
Under the present negligible-mean-cross-reactivity model, the CSK level changes the class mean mainly on the target axis. Let $\boldsymbol{\tilde{q}}_i = [\tilde{q}_i^{(t)},\tilde{q}_i^{(o)}]^\top$ and assume class-conditional Gaussian statistics with shared covariance
\begin{equation}
  \Sigma = \begin{bmatrix}
    \sigma_t^2 & \rho_{\mathrm{cc}}\sigma_t\sigma_o \\
    \rho_{\mathrm{cc}}\sigma_t\sigma_o & \sigma_o^2
  \end{bmatrix},
  \label{eq:csk_cov}
\end{equation}
and mean difference $\Delta\boldsymbol{\mu} = [\Delta_t,0]^\top$. The equal-covariance maximum-likelihood rule reduces to the linear discriminant $(\Sigma^{-1}\Delta\boldsymbol{\mu})^\top\boldsymbol{\tilde{q}}_i$. Since
\begin{equation}
  \Sigma^{-1}\Delta\boldsymbol{\mu}
  = \frac{\Delta_t}{\sigma_t^2(1-\rho_{\mathrm{cc}}^2)}
  \begin{bmatrix}
    1 \\
    -\rho_{\mathrm{cc}}\sigma_t/\sigma_o
  \end{bmatrix},
  \label{eq:csk_w}
\end{equation}
the discriminant is, up to an irrelevant positive scale factor,
\begin{equation}
  (\Sigma^{-1}\Delta\boldsymbol{\mu})^\top\boldsymbol{\tilde{q}}_i
  \propto \frac{\tilde{q}_i^{(t)}}{\sigma_t}
    - \rho_{\mathrm{cc}}\frac{\tilde{q}_i^{(o)}}{\sigma_o}.
  \label{eq:csk_stat_ml}
\end{equation}
Equation~\eqref{eq:csk_stat} is therefore not an ad hoc combiner. It is the equal-covariance Gaussian discriminant (equivalently, an LDA-style combiner) specialized to a target-axis mean shift with residual cross-axis covariance.

\paragraph{CSK-4 and hybrid detectors}
For CSK-4, the receiver compares $\Gamma_{\mathrm{CSK},i}$ to three calibrated thresholds to decide the transmitted level. For the hybrid scheme, the receiver first decides molecule identity using \eqref{eq:gamma_mosk}, yielding an estimate $\hat{b}_{M,i}\in\{0,1\}$ of whether the symbol is \DA or \HT. It then decides the amplitude bit on the MoSK-selected axis using a single-axis control-referenced statistic,
\begin{equation}
  \Gamma_{\mathrm{Hyb},i}
  = \frac{\tilde{q}_i^{(t)}}{\sigma_t},
  \quad
  t=\begin{cases}
    \DA,& \hat{b}_{M,i}=0,\\
    \HT,& \hat{b}_{M,i}=1,
  \end{cases}
  \label{eq:hyb_amp_stat}
\end{equation}
followed by comparison to a calibrated binary threshold $\tau_{\mathrm{Hyb}}^{(t)}$ with a comparator direction consistent with the sign of $q_{\mathrm{eff},t}$. Unlike CSK-4, which leverages both axes through \eqref{eq:csk_stat}, the hybrid amplitude detector uses only the selected axis to avoid coupling the amplitude decision to the non-selected species channel.

\subsection{Transport in Restricted Extracellular Space}

We model molecular transport in restricted extracellular space with volume fraction $\alpha$ and tortuosity $\lambda$, and we include first-order clearance with rate $k_{\mathrm{clear}}$. For a point impulse at distance $r$, the Green's function for number density is~\cite{sykova2008diffusion,nicholson1998diffusion}
\begin{equation}
  g(r,t)
  = \frac{1}{\alpha\bigl(4\pi D_{\mathrm{eff}} t\bigr)^{3/2}}
    \exp\!\left(
      -\frac{r^2}{4D_{\mathrm{eff}} t}
      - k_{\mathrm{clear}} t
    \right)\mathbbm{1}\{t>0\},
  \label{eq:greens}
\end{equation}
where $D_{\mathrm{eff}} = D/\lambda^2$~\cite{sykova2008diffusion}.

For an emission rate $s(t)$ in molecules per second, the resulting number density at the receiver location is the causal convolution
\begin{equation}
  c_n(r,t) = \int_{0}^{t} s(\tau)\,g\bigl(r,t-\tau\bigr)\,\mathrm{d}\tau,
  \label{eq:conv_num}
\end{equation}
with units molecules per cubic meter. Since the binding kinetics are parameterized in molar units, we convert number density to molar concentration via
\begin{equation}
  c(r,t) = \frac{c_n(r,t)}{1000\,N_A},
  \label{eq:conv_molar}
\end{equation}
where $N_A$ is Avogadro's constant and $1~\mathrm{M}=1000~\mathrm{mol}/\mathrm{m}^3$.

We use a finite rectangular burst of duration $T_{\mathrm{rel}}$ and total molecule budget $N_m$ per symbol to approximate an impulsive release relative to the symbol interval ($T_{\mathrm{rel}}\ll T_s$). ISI is captured by superposing contributions from previous symbols. For analytical insight, let $s\in\{\DA,\HT\}$ denote the transmitted species, let $\bar{N}_{m,i}^{(s)}$ be the nominal mean emission count selected by the modulation alphabet during symbol $i$, and let the realized emitted count satisfy $\tilde{N}_{m,i}^{(s)}\sim\mathrm{Poisson}(\bar{N}_{m,i}^{(s)})$. We then define window coefficients in molar units
\begin{equation}
  h_k(r,T_s,W)
  = \frac{1}{W}\int_{T_s-W}^{T_s} \frac{g\bigl(r,t+kT_s\bigr)}{1000\,N_A}\,\mathrm{d}t,\quad k=0,1,\ldots,
  \label{eq:hk}
\end{equation}
which quantify the contribution of the symbol sent $k$ intervals earlier to the window-averaged molar concentration in the current decision window. For the species-specific transport path associated with $s$, the corresponding window-averaged concentration can be summarized as
\begin{equation}
  \bar{c}_i^{(s)}
  = \sum_{k\ge 0} \tilde{N}_{m,i-k}^{(s)}\, h_k(r_s,T_s,W).
  \label{eq:windowed_conc}
\end{equation}
Under the negligible-cross-reactivity assumption used below, the matching selective channel responds only to its own species-specific concentration trajectory.

\noindent In ISI-enabled simulations we truncate the diffusion superposition in \eqref{eq:windowed_conc} to a finite history length $K_{\mathrm{ISI}}$. Using the coefficients in \eqref{eq:hk}, $K_{\mathrm{ISI}}$ is chosen as the smallest integer $K\ge 2$ such that
\begin{equation}
  \sum_{k=K}^{\infty} h_k(r,T_s,W) \le \varepsilon\, h_0(r,T_s,W),\quad \varepsilon=10^{-3},
  \label{eq:isi_memory_rule}
\end{equation}
and we cap $K_{\mathrm{ISI}}\le K_{\max}$ with $K_{\max}=60$.

\subsection{Aptamer Binding Kinetics}

Each functionalized gate presents $N_{\mathrm{apt}}$ binding sites for its target neuromodulator. Binding follows Langmuir kinetics driven by the local molar concentration $c(t)$ at the aptamer layer~\cite{langmuir1918adsorption}. The transport model produces a number density $c_n(t)$ in molecules per cubic meter via \eqref{eq:conv_num}, which is converted to molar concentration by \eqref{eq:conv_molar}. The binding dynamics are
\begin{equation}
  \frac{\mathrm{d}N_b(t)}{\mathrm{d}t}
  = k_{\mathrm{on}}^{\mathrm{eff}}\,c(t)\,\bigl(N_{\mathrm{apt}} - N_b(t)\bigr)
    - k_{\mathrm{off}}\,N_b(t),
  \label{eq:binding_ode}
\end{equation}
where $k_{\mathrm{off}}$ is the off-rate, and $k_{\mathrm{on}}^{\mathrm{eff}} = k_{\mathrm{on}}/(1+\mathrm{Da})$ is an optional low-order transport-limitation heuristic rather than a first-principles reduction. Here $\mathrm{Da}$ denotes an effective Damk\"ohler number, i.e., the ratio of a characteristic binding timescale to a characteristic mass-transfer timescale, used only to soften the on-rate when transport limitation becomes significant~\cite{lubken2022dynamic}.

Unless otherwise stated, we set $\mathrm{Da}=0$, which corresponds to reaction-limited binding with $k_{\mathrm{on}}^{\mathrm{eff}}=k_{\mathrm{on}}$. The formulation supports $\mathrm{Da}>0$ when a transport-limited regime is justified by independent characterization of mass-transfer conditions. In the time-domain Monte Carlo, stochastic occupancy is represented by a discrete-time birth-death approximation consistent with \eqref{eq:binding_ode}, rather than an exact continuous-time chemical master-equation simulation. The \CTRL channel has $k_{\mathrm{on}}=k_{\mathrm{off}}=0$ and therefore no specific binding.

\subsection{OECT Transduction and Correlated Electrical Noise}

Binding-induced gating is transduced into drain current through the OECT transconductance under a quasi-static small-signal approximation in which the internal OECT ionic/RC dynamics are faster than the symbol-level decision window, or are absorbed into effective parameters $g_m$ and $C_{\mathrm{tot}}$~\cite{rivnay2015thickness,rivnay2018oect,friedlein2018physics,hanzaee2025differential}. Reported PEDOT:PSS OECTs exhibit peak transconductance in the few-mS range at microscale geometries. For example, Khodagholy \emph{et al.} report a typical peak $g_m=\SI{2.7}{mS}$ (and $\approx\SI{4}{mS}$ for their best device)~\cite{khodagholy2013high}. Rivnay \emph{et al.} further show that, at fixed bias in the saturation regime, $g_m$ increases approximately linearly with $W d/L$ (and with $\mu C^\ast$), enabling larger $g_m$ through geometric and thickness optimization~\cite{rivnay2015thickness}. We therefore use $g_m=\SI{5}{mS}$ as an optimistic but literature-grounded high-performance operating point in our simulations. For channel $X$,
\begin{equation}
  i_{\mathrm{sig},X}(t)
  = g_m \Delta V_{g,X}(t)
  = g_m \frac{q^{(\mathrm{eff})}_X e}{C_{\mathrm{tot}}} N_b^{(X)}(t),
  \label{eq:oect_signal}
\end{equation}
\noindent Here $q^{(\mathrm{eff})}_X$ is a signed, dimensionless effective charge factor (in units of the elementary charge $e$) that captures the effective gate-voltage perturbation per binding event, including aptamer conformational charge redistribution and ionic screening under the chosen bias. Because this transduction is typically reported phenomenologically (e.g., as transfer-curve or current shifts) rather than as a unique ``charge per binding event,'' we treat $q^{(\mathrm{eff})}_X$ as a lumped device--chemistry coupling coefficient. Its sign can differ between aptamers (here \DA versus \HT) due to opposite binding-induced aptamer reorientation reported for dopamine and serotonin aptamer FETs, which yields opposite-signed $I_{DS}$ responses~\cite{nakatsuka2018aptamer}.

The magnitude $|q^{(\mathrm{eff})}_X|$ is not directly specified by a unique experimental observable and is therefore treated as an assumed lumped coupling parameter. Unless otherwise stated, we set $|q^{(\mathrm{eff})}_X|=0.35$ (Table~\ref{tab:sim-params}).

For PEDOT:PSS OECTs, the relevant gate/channel coupling is commonly described by a volumetric capacitance $C^\ast$ rather than a planar dielectric. Rivnay \emph{et al.} report $C^\ast\approx\SI{39}{F\,cm^{-3}}$ and extract a channel capacitance $C\approx\SI{82.2}{nF}$ for a device with $W=L=\SI{50}{\micro m}$ and $d=\SI{500}{nm}$, illustrating that tens-of-nF effective capacitances are typical at microscale dimensions~\cite{rivnay2015thickness}. We therefore adopt $C_{\mathrm{tot}}=\SI{50}{nF}$ as a representative effective capacitance in Table~\ref{tab:sim-params}.

The measured current is
\begin{equation}
  i_X(t) = i_{\mathrm{sig},X}(t) + i_{\mathrm{n},X}(t).
\end{equation}

We model additive electrical noise as the sum of thermal (white), flicker ($1/f$), and drift ($1/f^2$) components with effective single-sided current power spectral densities~\cite{johnson1928thermal,nyquist1928thermal,hooge1969noise,stoop2017noise,polyravas2019overlap,polyravas2020thickness_noise,jakobson2000isfet}
\begin{align}
  S_{\mathrm{th}}(f)   &= \frac{4k_{\!B}T}{R_{\mathrm{ch}}}, \label{eq:psd_th} \\
  S_{1/f}(f)           &= \frac{K_f I_{\mathrm{DC}}^{2}}{f}, \label{eq:psd_flicker} \\
  S_{\mathrm{drift}}(f)&= \frac{K_{\mathrm{drift}} I_{\mathrm{DC}}^{2}}{f^{2}}. \label{eq:psd_drift}
\end{align}
In PEDOT:PSS OECTs, measured drain-current spectra commonly exhibit $1/f$ behavior at low frequencies (typically $\lesssim$100~Hz) and are often interpreted using a charge-noise (gate-referred) model $S_{I_d}=g_m^2 S_{V_g}$~\cite{stoop2017noise,polyravas2019overlap,polyravas2020thickness_noise}.
We use a Hooge-form parameterization for convenience (via $K_f=\alpha_H/N_c$) and choose $\alpha_H$ (Table~\ref{tab:sim-params}) to represent an optimistic, low-noise regime consistent with reported gate-referred noise metrics for thick-channel PEDOT:PSS OECTs~\cite{polyravas2020thickness_noise}.
The $1/f^2$ ``drift'' component captures very-low-frequency baseline wander. Electrolyte-gated sensors have been reported to transition from $1/f$ to $1/f^2$ below a drift-dependent corner frequency~\cite{jakobson2000isfet,song2024drift}.
Accordingly, we set $K_{\mathrm{drift}}$ so that the implied corner frequency $f_c = K_{\mathrm{drift}}/K_f$ lies below the dominant decision-window frequencies (typically $f \gtrsim 1/W$), making drift a subdominant contributor over the symbol windows considered.
The shared electrolyte and gate-bias environment motivates modeling the low-frequency components as partially common-mode across channels, consistent with the use of differential/complementary OECT front-ends and drift-mitigation studies in electrolyte-gated sensing~\cite{hanzaee2025differential,uguz2024complementary,song2024drift}. In the model and simulations, flicker and drift components are generated as correlated triplets. We use $\rho$ to denote the pre-subtraction correlation between a selective channel and the matched \CTRL reference channel, while thermal noise is modeled as uncorrelated across channels.

Control subtraction implements differential referencing for amplitude decisions~\cite{hanzaee2025differential}. For each selective channel $t\in\{\DA,\HT\}$, we form the referenced waveform $\tilde{i}_t(t)=i_t(t)-i_{\CTRL}(t)$ prior to window integration. This is equivalent to the charge-domain difference in \eqref{eq:qdiff} when the same decision window is used. The subtraction leaves independent thermal noise largely unchanged except for a factor-of-two variance increase, while it strongly rejects common-mode flicker and drift that are correlated between the selective and control channels. After subtraction, the two referenced selective axes remain correlated. We denote this post-subtraction residual selective-axis correlation by $\rho_{\mathrm{cc}}$, to distinguish it from the pre-subtraction selective-\CTRL coefficient $\rho$. Under a perfectly symmetric matched-reference toy model with independent selective noises and a shared \CTRL term, $\rho_{\mathrm{cc}}=0.5$ is a natural baseline. Here we keep $\rho_{\mathrm{cc}}$ as a scenario-dependent nuisance parameter to absorb mismatch, unequal variances, and filtering differences. Unless otherwise noted, we use assumed baseline values $\rho=0.9$ and $\rho_{\mathrm{cc}}=0.5$ (Table~\ref{tab:sim-params}), intended to represent strong pre-subtraction common-mode drift and moderate residual correlation after subtraction. These coefficients are therefore scenario-dependent nuisance parameters rather than empirically fixed constants.

Integrating \eqref{eq:oect_signal} over the decision window links molecular binding to the mean charge statistic. For channel $X$,
\begin{equation}
  \mu_i^{(X)} \triangleq \mathbb{E}\!\left[q_i^{(X)}\right]
  = g_m \frac{q^{(\mathrm{eff})}_X e}{C_{\mathrm{tot}}}
    \int_{T_s-W}^{T_s} \mathbb{E}\!\left[N_b^{(X)}(t_i+t)\right]\,\mathrm{d}t.
  \label{eq:mean_charge}
\end{equation}
Equivalently, with a local window variable $u\in[0,W]$, one may write $\mu_i^{(X)} = g_m \frac{q^{(\mathrm{eff})}_X e}{C_{\mathrm{tot}}}\int_{0}^{W}\mathbb{E}[N_b^{(X)}(t_i+T_s-W+u)]\,\mathrm{d}u$, which makes the tail anchoring explicit. For wide-sense stationary current noise, the variance of the integrated charge can be expressed through the noise PSD and the rectangular-window transfer function,
\begin{equation}
  \sigma_{q,X}^2
  = \int_{0}^{\infty} S_X(f)\left|\frac{\sin(\pi f W)}{\pi f}\right|^2 \mathrm{d}f,
  \label{eq:charge_var}
\end{equation}
where $S_X(f)$ is the sum of \eqref{eq:psd_th}--\eqref{eq:psd_drift}.

Equation~\eqref{eq:charge_var} is retained as the formal PSD-based reference relation for integrated-charge variance under rectangular windowing. To keep the analytical description aligned with the implemented detector normalization used by the simulator, however, we distinguish this reference relation from the surrogate normalization variances used operationally in the z-score detector. Specifically, with $f_{\min}=1/\max(W,\Delta t)$ and $f_{\max}=\min(B_{\mathrm{det}}, f_s/2)$ (plus a negligible numerical safeguard in code to enforce $f_{\max}>f_{\min}$), we define
\begin{align}
  \hat{\sigma}_{q,\mathrm{th}}^2 &= S_{\mathrm{th}} W, \label{eq:surrogate_noise_start}\\
  \hat{\sigma}_{q,1/f}^2 &= K_f I_{\mathrm{DC}}^{2}
  \ln\!\left(\frac{f_{\max}}{f_{\min}}\right) W, \\
  \hat{\sigma}_{q,\mathrm{drift}}^2 &= K_{\mathrm{drift}} I_{\mathrm{DC}}^{2}
  \left(\frac{1}{f_{\min}}-\frac{1}{f_{\max}}\right) W,
  \label{eq:surrogate_noise}
\end{align}
with $\hat{\sigma}_{q,\mathrm{lf}}^2 \triangleq \hat{\sigma}_{q,1/f}^2 + \hat{\sigma}_{q,\mathrm{drift}}^2$. In the baseline regime $W\gg \Delta t$, this reduces to $f_{\min}\approx 1/W$. These $\hat{\sigma}$ quantities are implementation-level detector normalizers rather than exact closed-form evaluations of \eqref{eq:charge_var}. Operationally, they collapse the rectangular-window weighting into the window-length factor $W$ used by the detector code. They are used only to scale the z-score statistics and to parameterize the dual-axis combiner in \eqref{eq:csk_stat}, while SER and LoD are estimated from time-domain Monte Carlo realizations of the full correlated noise synthesis.

Using $\hat{\sigma}_{q,\mathrm{lf}}^2$ from \eqref{eq:surrogate_noise}, the corresponding single-ended and control-referenced surrogate charge variances are approximated by
\begin{align}
  \hat{\sigma}_{q,\mathrm{single}}^2 &\approx \hat{\sigma}_{q,\mathrm{th}}^2 + \hat{\sigma}_{q,\mathrm{lf}}^2, \\
  \hat{\sigma}_{q,\mathrm{ref}}^2 &\approx 2\hat{\sigma}_{q,\mathrm{th}}^2 + 2(1-\rho)\hat{\sigma}_{q,\mathrm{lf}}^2.
  \label{eq:ref_var}
\end{align}
The factor of two on the thermal term follows from subtraction of independent channel noises, whereas the low-frequency term is reduced by the pre-subtraction common-mode correlation $\rho$. Since the \CTRL channel carries no intended analyte signal, control referencing is beneficial for amplitude detection when $\hat{\sigma}_{q,\mathrm{ref}}^2 < \hat{\sigma}_{q,\mathrm{single}}^2$, which yields
\begin{equation}
  \hat{\sigma}_{q,\mathrm{th}}^2 < (2\rho - 1)\hat{\sigma}_{q,\mathrm{lf}}^2
  \Longleftrightarrow
  \rho > \frac{1}{2}\left(1 + \frac{\hat{\sigma}_{q,\mathrm{th}}^2}{\hat{\sigma}_{q,\mathrm{lf}}^2}\right).
  \label{eq:control_benefit}
\end{equation}
Equation~\eqref{eq:control_benefit} is approximate because it is derived from surrogate detector normalizers rather than from direct numerical evaluation of \eqref{eq:charge_var}, and because the full receiver also includes finite-window integration, threshold calibration, and multichannel Monte Carlo realizations. Nevertheless, it captures the observed regime rule. When low-frequency nuisance dominates, the crossover tends toward $\rho>0.5$. When thermal noise is comparable or larger, subtraction can be neutral or harmful. The residual correlation $\rho_{\mathrm{cc}}$ between the two control-referenced selective axes is accounted for in \eqref{eq:csk_stat}.

\subsection{Error Metrics and Limit of Detection}

We use symbol error rate (SER) and limit of detection (LoD) as the primary performance metrics. When a scalar decision statistic is approximately Gaussian under two hypotheses with equal priors and equal variances $\sigma_0=\sigma_1=\sigma$, the binary error probability is
\begin{align}
  \mathrm{SER}
  &\approx Q\!\left(\frac{|\mu_1-\mu_0|}{2\sigma}\right)
  = \frac{1}{2}\,\mathrm{erfc}\!\left(
    \frac{|\mu_1-\mu_0|}{2\sqrt{2}\,\sigma}
  \right) \nonumber\\
  &= \frac{1}{2}\,\mathrm{erfc}\!\left(\sqrt{\mathrm{SNR}/4}\right),
  \label{eq:gaussian_ser}
\end{align}
where $Q(\cdot)$ is the standard Gaussian tail function, $\mathrm{SNR} = (\mu_1-\mu_0)^2/(\sigma_0^2+\sigma_1^2)$, and the final equality uses $\sigma_0=\sigma_1=\sigma$. For unequal variances, the optimal threshold shifts and \eqref{eq:gaussian_ser} should be interpreted only as an equal-variance bracketing formula. This approximation is used only for operating-point bracketing and intuition. All SER and LoD values reported in Section~\ref{sec:results} are obtained from Monte Carlo simulation of the full time-domain receiver.

For a fixed separation $r$, the LoD at target error rate $\varepsilon$ is defined as
\begin{equation}
  \mathrm{LoD}(r)
  = \min\left\{N_m : \mathrm{SER}(N_m,r)\leq \varepsilon\right\},\quad \varepsilon = 0.01.
  \label{eq:lod_def}
\end{equation}
We adopt $\varepsilon=10^{-2}$ as a reference receiver target throughout the paper.

\subsection{Summary of Modeling Assumptions}

The analytical model and numerical evaluation share the following assumptions.

\begin{itemize}
  \item Transport follows the three-dimensional Green's function in \eqref{eq:greens} with clearance and an effective restricted medium characterized by $\alpha$ and $\lambda$. Concentrations are computed as number density and converted to molar units for binding via \eqref{eq:conv_molar}.
  \item Release uses a finite rectangular burst of duration $T_{\mathrm{rel}}$ per symbol. ISI is modeled by superposition of past-symbol concentration tails, and it can be summarized through window coefficients $h_k(r,T_s,W)$. The transport stage is treated at the mean-field Green's-function level. Randomness enters through Poisson emission counts and downstream stochastic binding. We do not introduce a separate independent diffusion-arrival noise term beyond that receiver-level abstraction.
  \item Aptamer binding follows Langmuir kinetics as in \eqref{eq:binding_ode}. A Damk\"ohler-inspired transport-limitation factor is included in the framework through $k_{\mathrm{on}}^{\mathrm{eff}}$, and the baseline sweeps use reaction-limited binding with $\mathrm{Da}=0$. In Monte Carlo, stochastic occupancy is implemented as a discrete-time birth-death approximation consistent with this ODE rather than as an exact continuous-time SSA.
  \item We assume negligible aptamer cross-reactivity and non-specific binding. Under \DA transmission, the \HT selective gate has zero mean binding response and vice versa, while the \CTRL gate has no specific binding. This is consistent with reports of high small-molecule aptamer selectivity between monoamine neurotransmitters~\cite{nakatsuka2018aptamer}. Any residual coupling is treated as noise, and the mean leakage fraction in the dual-channel CSK statistic is set to zero unless stated otherwise.
  \item OECT transduction uses the quasi-static small-signal mapping in \eqref{eq:oect_signal} with the device parameters in Tables~\ref{tab:device_geom} and~\ref{tab:sim-params}.
  \item Electrical noise includes thermal, flicker, and drift components. Low-frequency components are correlated across channels due to shared bias, while thermal noise is treated as uncorrelated.
  \item MoSK uses the sign-aware statistic in \eqref{eq:gamma_mosk}. CSK-4 uses control-referenced charges in \eqref{eq:qdiff} and the dual-channel contrast in \eqref{eq:csk_stat}. The hybrid scheme combines MoSK for molecule identity with a single-axis control-referenced amplitude statistic in \eqref{eq:hyb_amp_stat}.
\end{itemize}

\section{Simulation Framework}
\label{sec:simulation}

Numerical results are obtained from a custom Monte Carlo simulator implementing the model in Section~\ref{sec:analytical}. Unless a figure explicitly sweeps a parameter, Section~\ref{sec:results} uses the baseline values in Table~\ref{tab:sim-params}. The simulator advances transport, stochastic binding, and OECT transduction at fixed time step $\Delta t=\SI{0.01}{s}$ and forms decision statistics using the tail-anchored window fraction $\eta=0.6$ in \eqref{eq:window_fraction}. Unless otherwise stated, decision statistics are evaluated in the variance-normalized (z-score) domain (detector mode ``zscore''). The normalization uses the surrogate detector normalizers in \eqref{eq:surrogate_noise_start}--\eqref{eq:control_benefit}, rather than repeated numerical evaluation of \eqref{eq:charge_var}, so that thresholds are dimensionless and comparable across operating points. Baseline SER and LoD sweeps disable diffusion-tail ISI to isolate noise-limited receiver performance. ISI-enabled sweeps reuse the same parameters and incorporate concentration superposition across past symbols and stateful aptamer occupancy carryover. When ISI is enabled, the number of past symbols included in the concentration superposition is chosen using the window-coefficient rule in \eqref{eq:isi_memory_rule}, and it is capped at 60 symbols.

For each operating point $(N_m,r)$, we estimate the SER by simulating a random symbol sequence of length 2000 symbols per Monte Carlo seed and averaging across independent seeds. We use a minimum of 8 seeds per point and add seeds until the Wilson 95\% confidence interval~\cite{wilson1927probable} half-width satisfies $\Delta\mathrm{SER}\le 4\times 10^{-3}$ near $\mathrm{SER}=0.01$, or until a maximum of 50 seeds is reached.

For LoD searches, we first bracket $N_m$ using the equal-variance Gaussian SER approximation in \eqref{eq:gaussian_ser} to avoid unnecessary Monte Carlo evaluations, then refine the smallest $N_m$ meeting \eqref{eq:lod_def} via distance-aware Monte Carlo validation.

Unless otherwise specified, the organoid-to-gate separation is swept over $r\in\{25,30,35,40,45,60,75,90,110,130\}\,\si{\micro\meter}$. Symbol periods are selected per distance using \eqref{eq:dynamic_ts} with guard factor $g=0.15$ and minimum symbol period $T_{\min}=\SI{5}{s}$. The random seed is fixed per run for reproducibility while each Monte Carlo seed uses independent stochastic realizations of molecular shot noise, binding, and device noise.

\subsection{Parameterization}

Table~\ref{tab:sim-params} summarizes the numerical values used throughout the simulations. It also distinguishes literature-grounded quantities from scenario baselines and assumed lumped parameters. The restricted-space transport parameters $(\alpha,\lambda)$, aqueous diffusivities, aptamer affinities, and representative OECT operating point are grounded in prior studies~\cite{sykova2008diffusion,khodagholy2013high,rivnay2015thickness,hanzaee2025differential}. By contrast, $k_{\mathrm{clear}}$, $\rho$, and $\rho_{\mathrm{cc}}$ are scenario parameters chosen to study slow-clearance, common-mode-limited operation, and $q_{\mathrm{eff}}$ is an assumed lumped electrostatic coupling factor. The absolute performance values in Section~\ref{sec:results} should therefore be read as scenario-based receiver forecasts. The more transferable contribution is the regime-dependent design rule for when matched control referencing helps the Hybrid amplitude branch.

\begin{table*}[!t]
  \centering
  \caption{Baseline parameters, numerical settings, and anchoring used throughout the simulations.}
  \label{tab:sim-params}
  \footnotesize
  \setlength{\tabcolsep}{4pt}
  \renewcommand{\arraystretch}{1.05}
  \begin{tabularx}{\textwidth}{@{}p{0.17\textwidth}p{0.29\textwidth}X@{}}
  \toprule
  Quantity & Baseline setting & Status and anchoring \\
  \midrule
  Restricted medium & $\alpha=0.20$, $\lambda=1.6$, $D_{\mathrm{eff}}=D/\lambda^2$, concentration scaled by $1/\alpha$, $k_{\mathrm{clear}}=\SI{0.01}{s^{-1}}$ & $\alpha$ and $\lambda$ follow brain extracellular-space literature~\cite{sykova2008diffusion}. The slow-clearance value is a conservative scenario baseline for diffusion-dominated ISI, not an organoid-validated nominal. Commonly used striatal values imply much faster uptake, $V_{\max}/K_m\approx 25~\mathrm{s}^{-1}$~\cite{wightman1988dopamine,walters2014restricted,hoffman2016transport}. \\
  Aqueous diffusion & $D_{\DA}=4.9\times 10^{-10}~\mathrm{m}^2/\mathrm{s}$, $D_{\HT}=5.3\times 10^{-10}~\mathrm{m}^2/\mathrm{s}$ & Literature-grounded free-diffusion values for DA and 5-HT~\cite{colombo2016dopamine,gerhardt1982diffusion}. \\
  Release profile & Rectangular burst with $T_{\mathrm{rel}}=\SI{10}{ms}$ & Assumed numerical impulse with $T_{\mathrm{rel}}\ll T_s$. \\
  Binding kinetics & $k_{\mathrm{on}}=10^{5}~\mathrm{M}^{-1}\mathrm{s}^{-1}$, $k_{\mathrm{off}}^{\DA}=0.015~\mathrm{s}^{-1}$, $k_{\mathrm{off}}^{\HT}=0.003~\mathrm{s}^{-1}$ & $k_{\mathrm{on}}$ follows reported small-molecule aptamer kinetics~\cite{chang2014kinetic}. Off-rates are derived from reported aptamer affinities, $K_d=\SI{150}{nM}$ for DA and $K_d=\SI{30}{nM}$ for 5-HT~\cite{nakatsuka2018aptamer}. \\
  Aptamer sites & $N_{\mathrm{apt}}=2\times10^8$ effective sites per selective gate & Conservative effective site count consistent with reported ssDNA surface coverages on Au~\cite{steel1998electrochemical}. \\
  OECT operating point & $g_m=\SI{5}{mS}$, $C_{\mathrm{tot}}=\SI{50}{nF}$, $R_{\mathrm{ch}}=\SI{500}{\ohm}$, $I_{\mathrm{DC}}=\SI{100}{\micro A}$, $V_g=\SI{-0.2}{V}$ & Representative high-performance OECT regime from the literature~\cite{khodagholy2013high,rivnay2015thickness,rivnay2018oect,hanzaee2025differential}. \\
  Effective charge factor & $q_\mathrm{eff}^{\DA}=-0.35$, $q_\mathrm{eff}^{\HT}=+0.35$ & Assumed lumped per-binding coupling that captures conformational charge redistribution, ionic screening, and capacitive coupling. The sign choice is a modeling assumption motivated by opposite-signed dopamine and serotonin aptamer-FET $I_{DS}$ responses~\cite{nakatsuka2018aptamer}, not a validated property of the exact OECT stack analyzed here. \\
  Electrical noise & Thermal + flicker + drift, $\alpha_H=10^{-3}$, $N_c=4.5\times10^{11}$, $K_{\mathrm{drift}}=10^{-16}~\mathrm{Hz}$, $B_{\mathrm{det}}=\SI{100}{Hz}$ & Representative low-noise OECT setting used for receiver study~\cite{johnson1928thermal,nyquist1928thermal,hooge1969noise,stoop2017noise,polyravas2019overlap,polyravas2020thickness_noise,jakobson2000isfet}. \\
  Cross-channel correlation & $\rho=0.9$ before subtraction, $\rho_{\mathrm{cc}}=0.5$ after subtraction & Scenario nuisance parameters representing strong shared drift before subtraction and moderate residual correlation after subtraction. \\
  Timing and ISI & $\Delta t=\SI{0.01}{s}$, $\eta=0.6$, $g=0.15$, $T_s$ from \eqref{eq:dynamic_ts} & Numerical setting and distance-dependent symbol-period policy. \\
  SER estimation & 2000 symbols/seed, 8 to 50 seeds, Wilson target $\Delta\mathrm{SER}\le 4\times10^{-3}$ near $\mathrm{SER}=0.01$ & Confidence-controlled Monte Carlo evaluation. \\
  LoD search & Bracketed search in $N_m$ plus Monte Carlo validation & Distance-aware validation schedule described in Section~\ref{sec:simulation}. \\
  \bottomrule
  \end{tabularx}
\end{table*}

\subsection{End-to-End Flow and Threshold Calibration}

For each symbol, the simulation generates a molecular release, computes concentration at each gate using the restricted-diffusion model with clearance and ISI superposition, simulates stochastic aptamer occupancy, maps occupancy to drain current using the OECT gain, adds tri-channel correlated device noise, and evaluates the appropriate decision statistic.

Detector thresholds for MoSK, CSK-4, and the hybrid amplitude bit are obtained from calibration sequences at the operating point. Calibration disables ISI so that thresholds reflect single-symbol decision statistics under the same noise and decision-window settings used in evaluation. Thresholds are computed as maximum-likelihood boundaries between adjacent levels. LoD at distance $r$ is the smallest $N_m$ such that $\mathrm{SER}(N_m,r)\le 0.01$ as in \eqref{eq:lod_def}. These per-operating-point thresholds make the baseline SER and LoD curves calibrated front-end benchmarks rather than fixed-threshold deployment forecasts.

\section{Numerical Results}
\label{sec:results}

Unless otherwise stated, all results in this section use the baseline parameters in Table~\ref{tab:sim-params} and are obtained from the Monte Carlo procedure in Section~\ref{sec:simulation}. Each plotted marker represents a direct numerical estimate at one operating point, and the solid lines are included only as visual guides. All curves are direct simulator outputs rather than post hoc fits. Baseline SER and LoD sweeps disable ISI to isolate the front-end and detection behavior, so they should be read as calibrated no-ISI front-end benchmarks under the scenario parameters of Table~\ref{tab:sim-params}. Fig.~\ref{fig:ts_sweep} explicitly re-enables ISI to study symbol-period selection.

\subsection{Baseline Hybrid Performance and Error Decomposition}

\begin{figure*}[!t]
  \centering
  \subfloat[Baseline SER at $r=\SI{45}{\micro\meter}$ and no ISI.]{
    \includegraphics[width=0.48\textwidth]{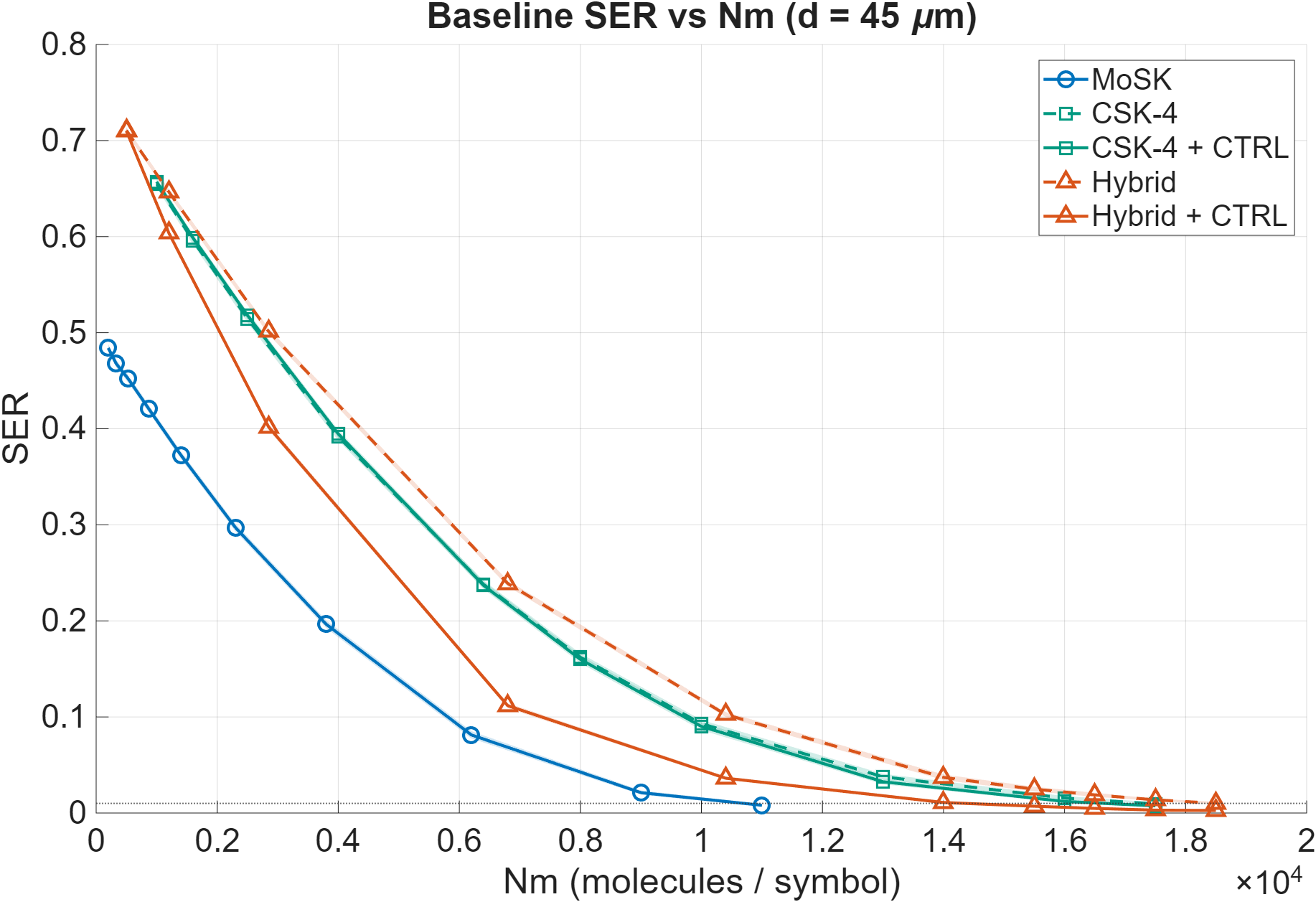}
    \label{fig:baseline_ser}
  }
  \hfill
  \subfloat[Hybrid decomposition into molecule-identity and amplitude-bit components.]{
    \includegraphics[width=0.48\textwidth]{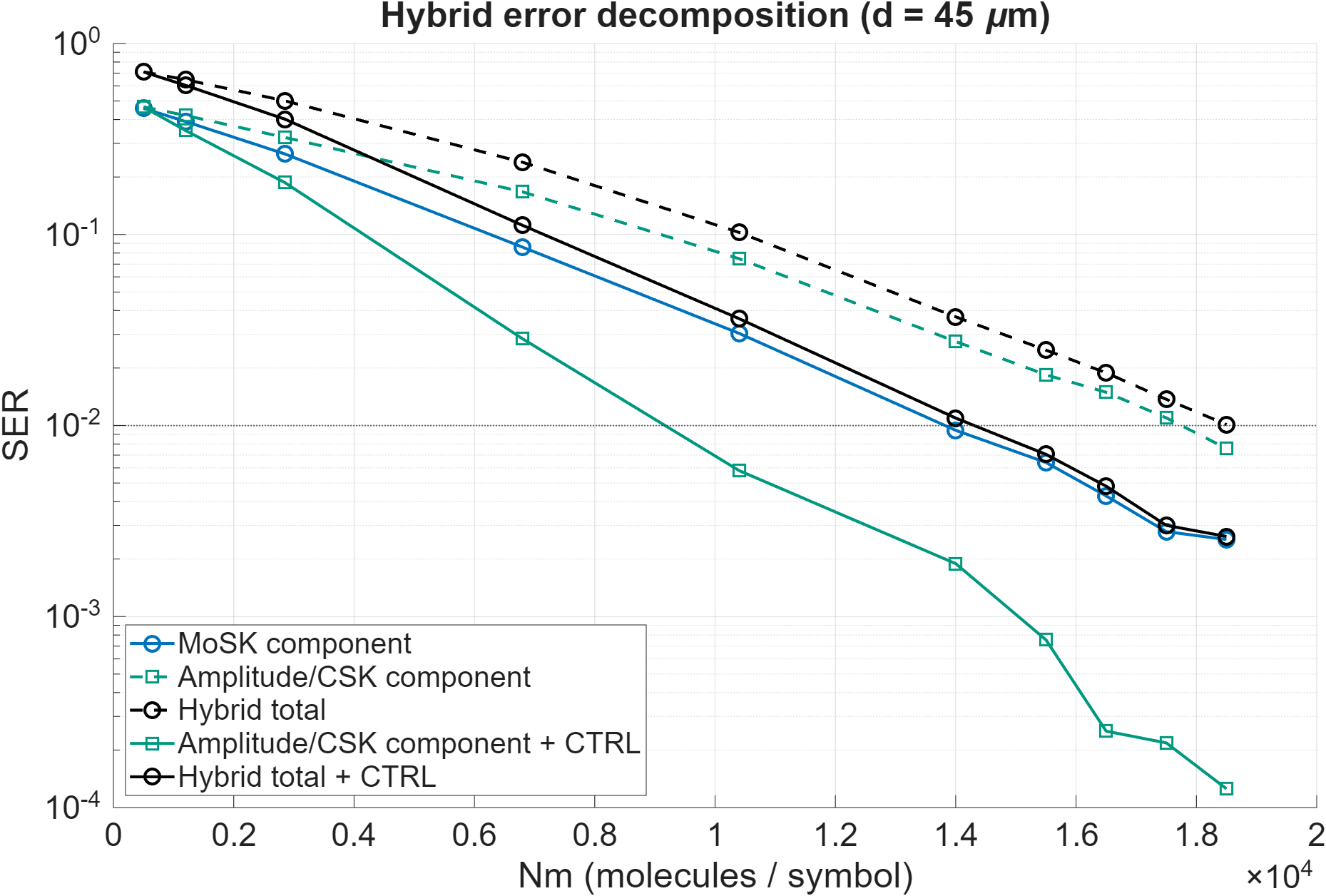}
    \label{fig:hybrid_decomp}
  }
  \caption{Nominal-spacing performance under the Table~\ref{tab:sim-params} baseline. MoSK does not use the control channel. CSK-4 and Hybrid are shown with and without control referencing. In the Hybrid receiver, the control channel primarily suppresses the amplitude-bit error component rather than the MoSK identity component. Lines are guides to the eye between simulated operating points.}
  \label{fig:baseline_results}
\end{figure*}

Figure~\ref{fig:baseline_results}(a) establishes the nominal operating point at $r=\SI{45}{\micro\meter}$. As expected, MoSK reaches low SER with the smallest molecule budget because it carries one bit in molecule identity only. The two-bit schemes require more molecules, but the key result is that Hybrid becomes competitive only when its amplitude decision is control referenced. Hybrid+CTRL first crosses the 1\% SER target at $N_m=1.55\times 10^4$ molecules/symbol, whereas Hybrid without CTRL remains above 1\% over the simulated range up to $1.85\times 10^4$. Under the same conditions, CSK-4+CTRL reaches the 1\% region at $1.75\times 10^4$ molecules/symbol. The tri-channel receiver therefore enlarges the baseline feasibility region of the intended two-bit Hybrid symbol rather than merely reproducing the MoSK baseline.

The mechanism is shown in Fig.~\ref{fig:baseline_results}(b). Without CTRL, the Hybrid SER is dominated by the amplitude bit, not by molecule identity. At $N_m=1.40\times 10^4$, the MoSK component is already below 1\% ($9.81\times 10^{-3}$), whereas the amplitude component is $2.73\times 10^{-2}$, yielding a total Hybrid SER of $3.71\times 10^{-2}$. Enabling CTRL changes the MoSK component only marginally ($9.06\times 10^{-3}$) but reduces the amplitude component to $1.88\times 10^{-3}$, which lowers the total Hybrid SER to $1.09\times 10^{-2}$. The 93.1\% reduction of the amplitude component at this operating point directly explains why the additional control channel is useful for Hybrid reception.

\subsection{Distance Dependence and Limit of Detection}

\begin{figure*}[!t]
  \centering
  \begin{minipage}[t]{0.31\textwidth}
    \centering
    \subfloat[LoD versus organoid-to-gate separation.\label{fig:lod_vs_distance}]{%
      \includegraphics[width=\linewidth]{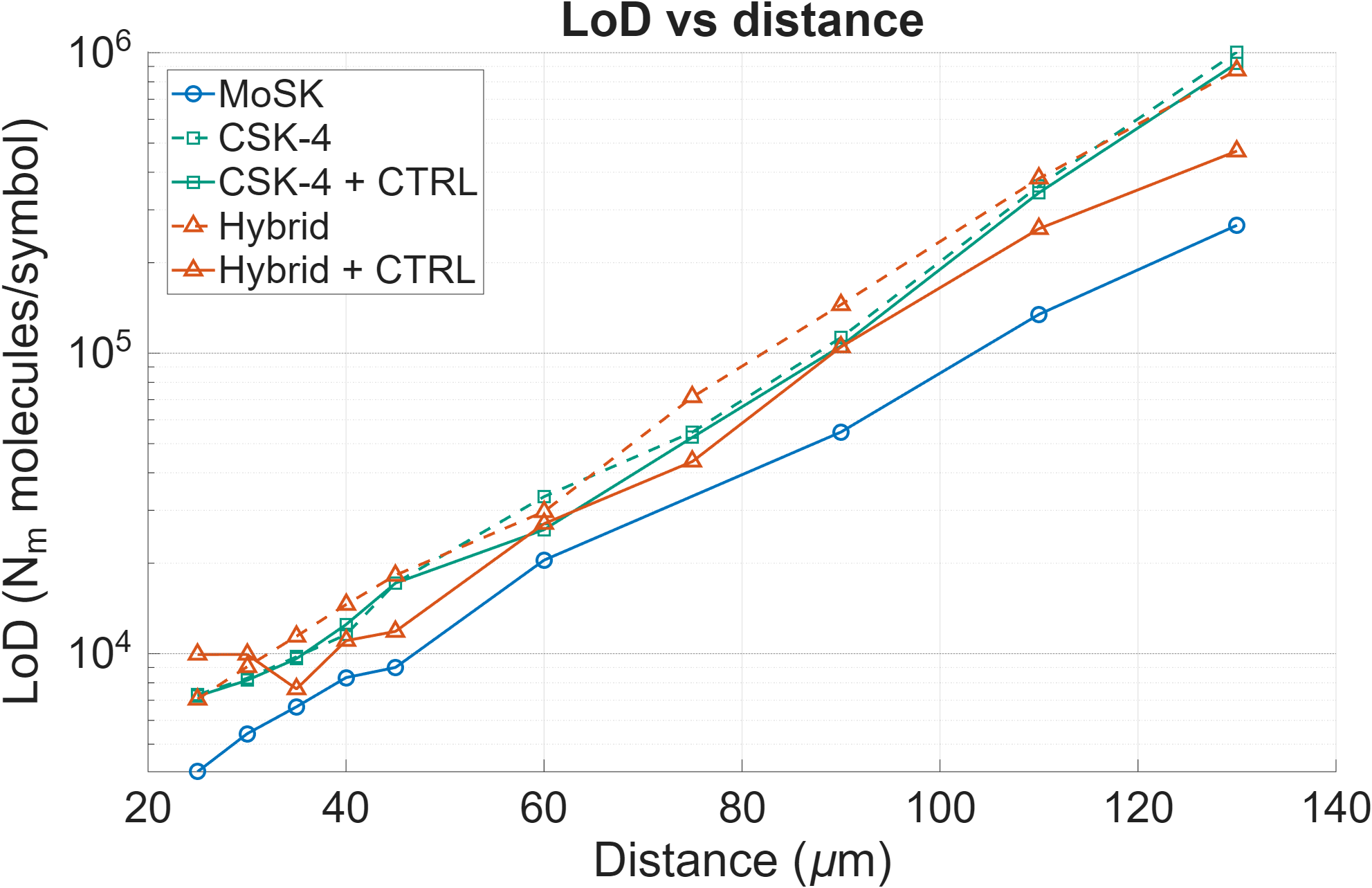}}
  \end{minipage}\hfill
  \begin{minipage}[t]{0.31\textwidth}
    \centering
    \subfloat[Hybrid control gain, defined as $\mathrm{LoD}_{\mathrm{off}}/\mathrm{LoD}_{\mathrm{on}}$.\label{fig:ctrl_gain_distance}]{%
      \includegraphics[width=\linewidth]{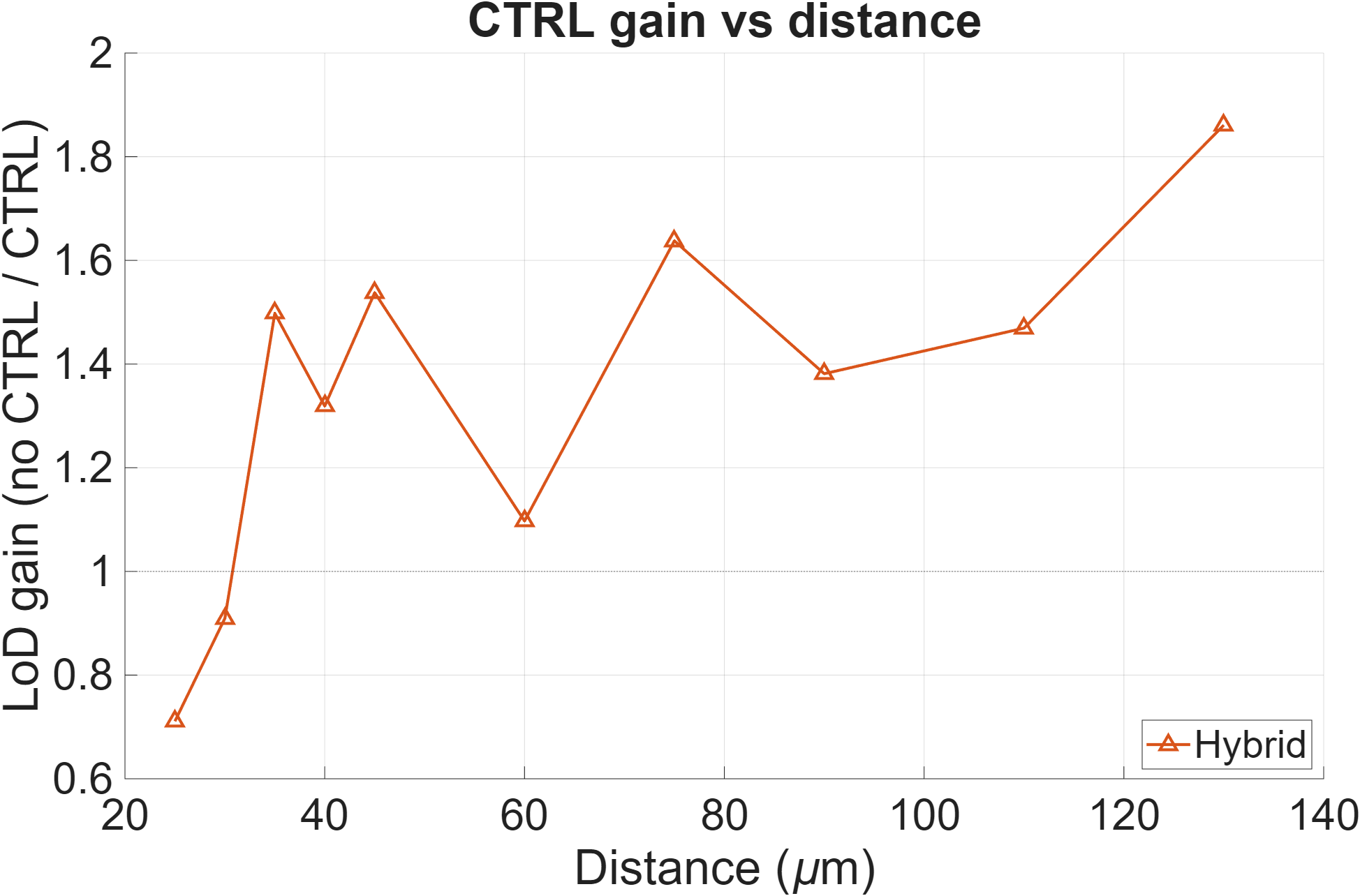}}
  \end{minipage}\hfill
  \begin{minipage}[t]{0.31\textwidth}
    \centering
    \subfloat[Hybrid SER versus distance at fixed $N_m=23972$ molecules/symbol.\label{fig:ser_distance_hybrid}]{%
      \includegraphics[width=\linewidth]{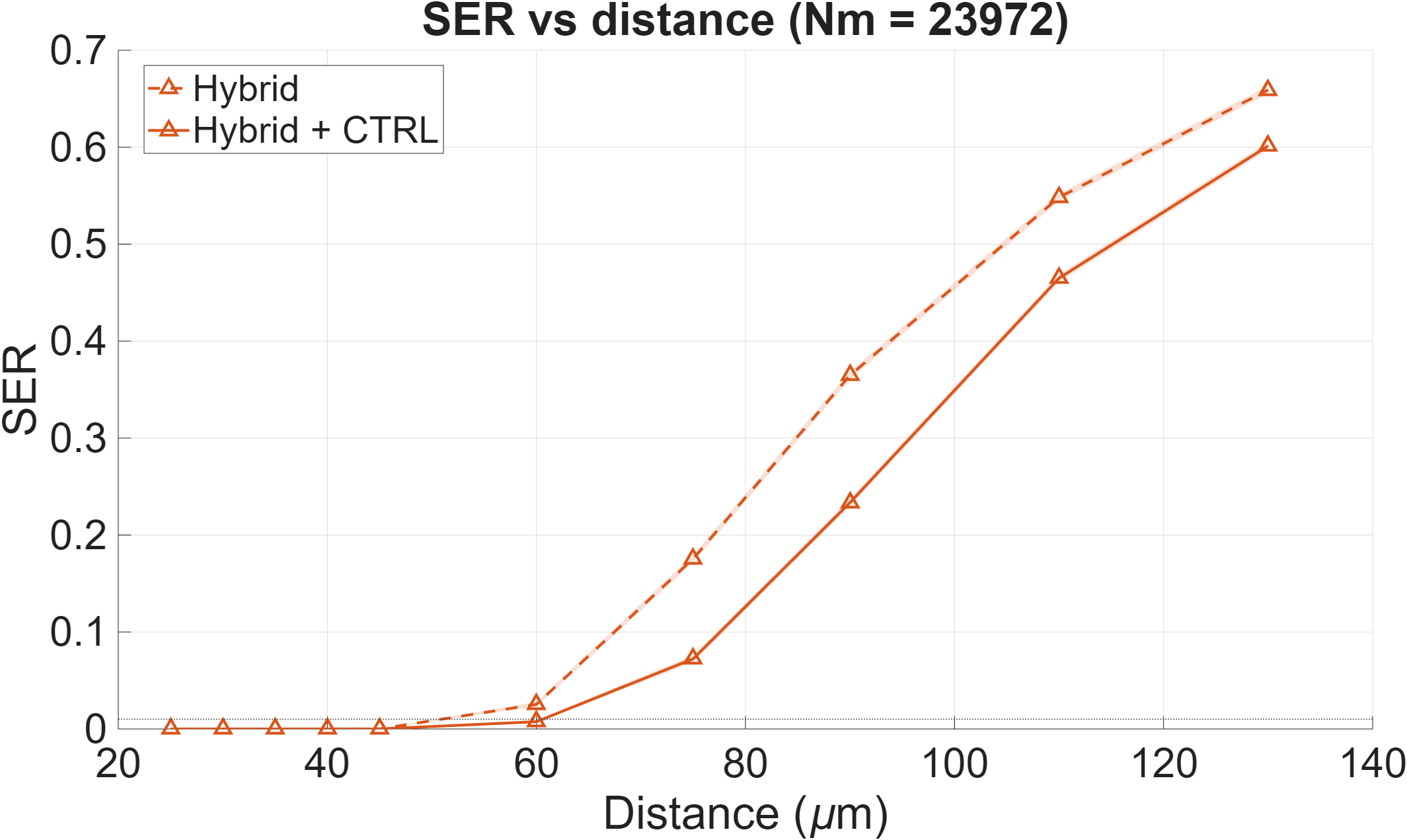}}
  \end{minipage}
  \caption{Distance-dependent behavior under the baseline device setting. The LoD is defined as the smallest molecule budget meeting $\mathrm{SER}\le 10^{-2}$. In (b), values above 1 indicate that control referencing lowers the Hybrid LoD.}
  \label{fig:distance_results}
\end{figure*}

Figure~\ref{fig:distance_results}(a) converts SER into the receiver-oriented LoD metric. MoSK retains the lowest absolute LoD because it transmits one bit/symbol, so the more relevant comparison is between the two-bit schemes. At $r=\SI{45}{\micro\meter}$, Hybrid+CTRL reaches an LoD of 11866 molecules/symbol compared with 17177 for CSK-4+CTRL, a 30.9\% reduction. At $r=\SI{130}{\micro\meter}$, the same comparison is 470791 versus 922484 molecules/symbol, i.e., a 49.0\% reduction. Relative to MoSK, Hybrid+CTRL at \SI{45}{\micro\meter} requires only 31.8\% more molecules while carrying twice the symbol payload.

The CTRL path is not uniformly beneficial, which is precisely why the control channel must be interpreted mechanistically rather than heuristically. For Hybrid, control referencing slightly worsens LoD at 25 and \SI{30}{\micro\meter} (9937 and 9944 molecules/symbol with CTRL versus 7064 and 9043 without CTRL), because subtraction adds uncorrelated reference noise when the signal is still strong. The crossover occurs between 30 and \SI{35}{\micro\meter}. Beyond \SI{35}{\micro\meter} the control gain remains greater than unity and rises to about 1.86 at \SI{130}{\micro\meter}, as summarized in Fig.~\ref{fig:distance_results}(b). Hybrid+CTRL is also not the minimum two-bit curve at every single distance point. At \SI{60}{\micro\meter}, for example, CSK-4+CTRL is marginally lower. Nevertheless, Hybrid+CTRL gives the strongest overall two-bit trend over most of the medium-to-long-distance regime in Fig.~\ref{fig:distance_results}(a).

Figure~\ref{fig:distance_results}(c) shows the same transition at fixed $N_m=23972$ molecules/symbol. Up to \SI{45}{\micro\meter} both Hybrid curves are near zero error, so the benefit of control referencing is small. Once the gap reaches 60--\SI{90}{\micro\meter}, the advantage becomes clear: the SER drops from $2.53\times 10^{-2}$ to $7.50\times 10^{-3}$ at \SI{60}{\micro\meter} and from $1.76\times 10^{-1}$ to $7.24\times 10^{-2}$ at \SI{75}{\micro\meter}. At the longest distances both curves degrade, but CTRL still preserves a lower SER. These results identify the distance range in which a matched control axis materially extends the useful operating envelope of the receiver.

\subsection{Symbol-Period Selection Under ISI}

\begin{figure*}[!t]
  \centering
  \includegraphics[width=0.72\textwidth]{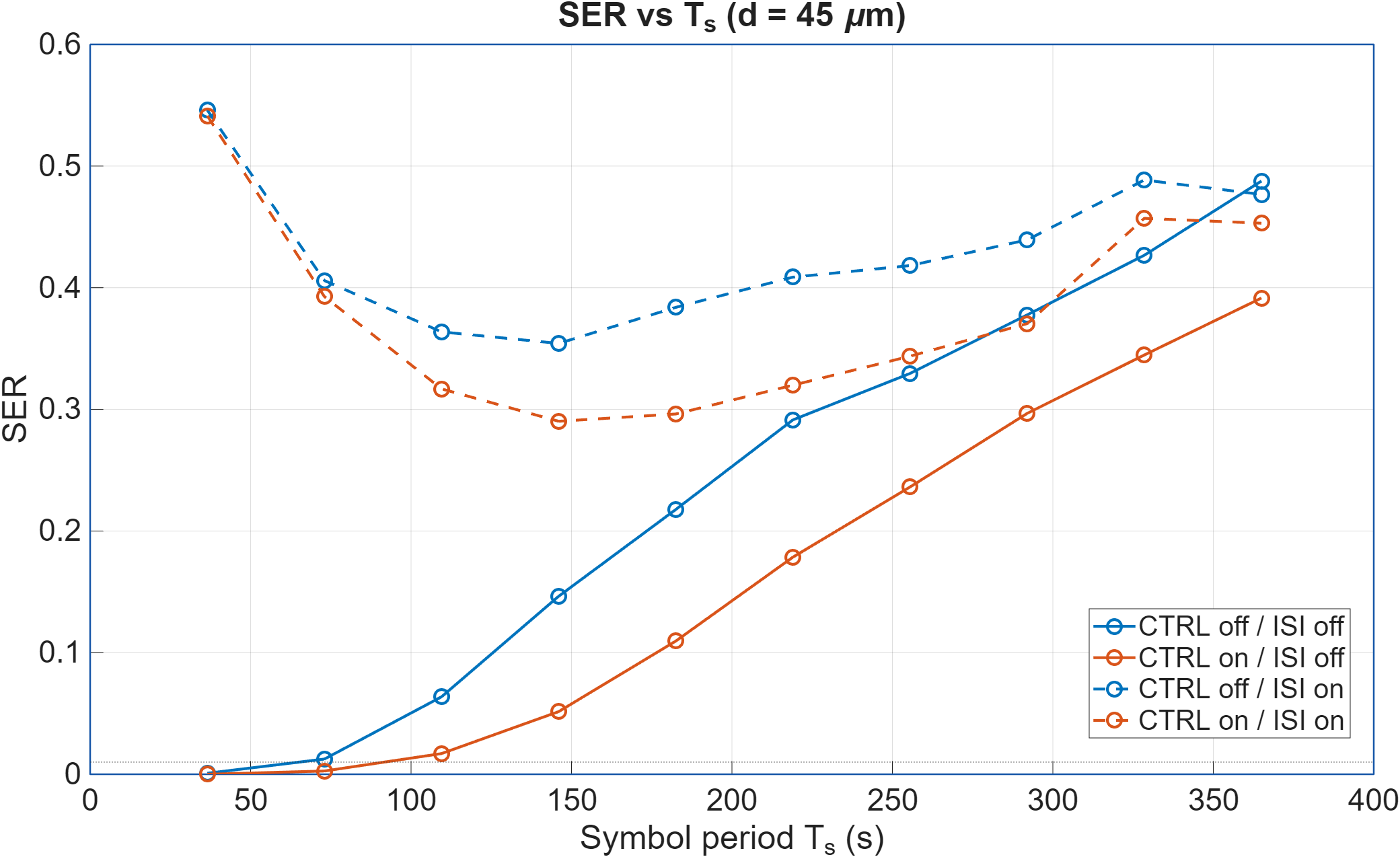}
  \caption{Hybrid SER versus symbol period at $r=\SI{45}{\micro\meter}$ and $N_m=2.0\times 10^4$, with and without ISI and with and without control referencing. The SER minimum shifts from the shortest tested $T_s$ under no-ISI conditions to an intermediate $T_s$ once molecular memory is enabled.}
  \label{fig:ts_sweep}
\end{figure*}

Figure~\ref{fig:ts_sweep} isolates the timing tradeoff at $r=\SI{45}{\micro\meter}$ and $N_m=2.0\times 10^4$. When ISI is disabled, the shortest tested symbol period ($T_s=\SI{36.5}{s}$) gives the lowest SER because the useful molecular response is concentrated into the strongest portion of the decision window. Once ISI is enabled, however, the optimum shifts to an intermediate symbol period. For Hybrid, the ISI-on SER falls from 0.546 at $T_s=\SI{36.5}{s}$ to 0.354 without CTRL and 0.290 with CTRL at $T_s\approx\SI{146}{s}$, and then rises again for longer periods. Short periods suffer from residual concentration and occupancy carryover, whereas overly long periods dilute the useful tail-window charge through clearance and binding relaxation.

Supporting MoSK and CSK-4 sweeps (not plotted separately here) show the same qualitative behavior: the ISI-optimal period remains near \SI{146}{s} for MoSK and shifts to about \SI{109.5}{s} for CSK-4 under the same baseline conditions. Hence, symbol-period selection is part of receiver design rather than a purely link-layer throughput choice. At the Hybrid optimum $T_s\approx\SI{146}{s}$, a 2-bit symbol corresponds to a raw signaling rate of about 49 bit/h before accounting for errors, so the present receiver should be interpreted as a slow-state-readout architecture rather than a high-rate MC link. The CTRL channel lowers the Hybrid SER across the ISI-limited region, but it does not remove the need for an intermediate $T_s$.

\subsection{OECT Device Co-Design and Feasible Operating Envelope}

\begin{figure*}[!t]
  \centering
  \subfloat[SER versus $g_m$ at fixed $C_{\mathrm{tot}}=\SI{50}{nF}$.]{
    \includegraphics[width=0.48\textwidth]{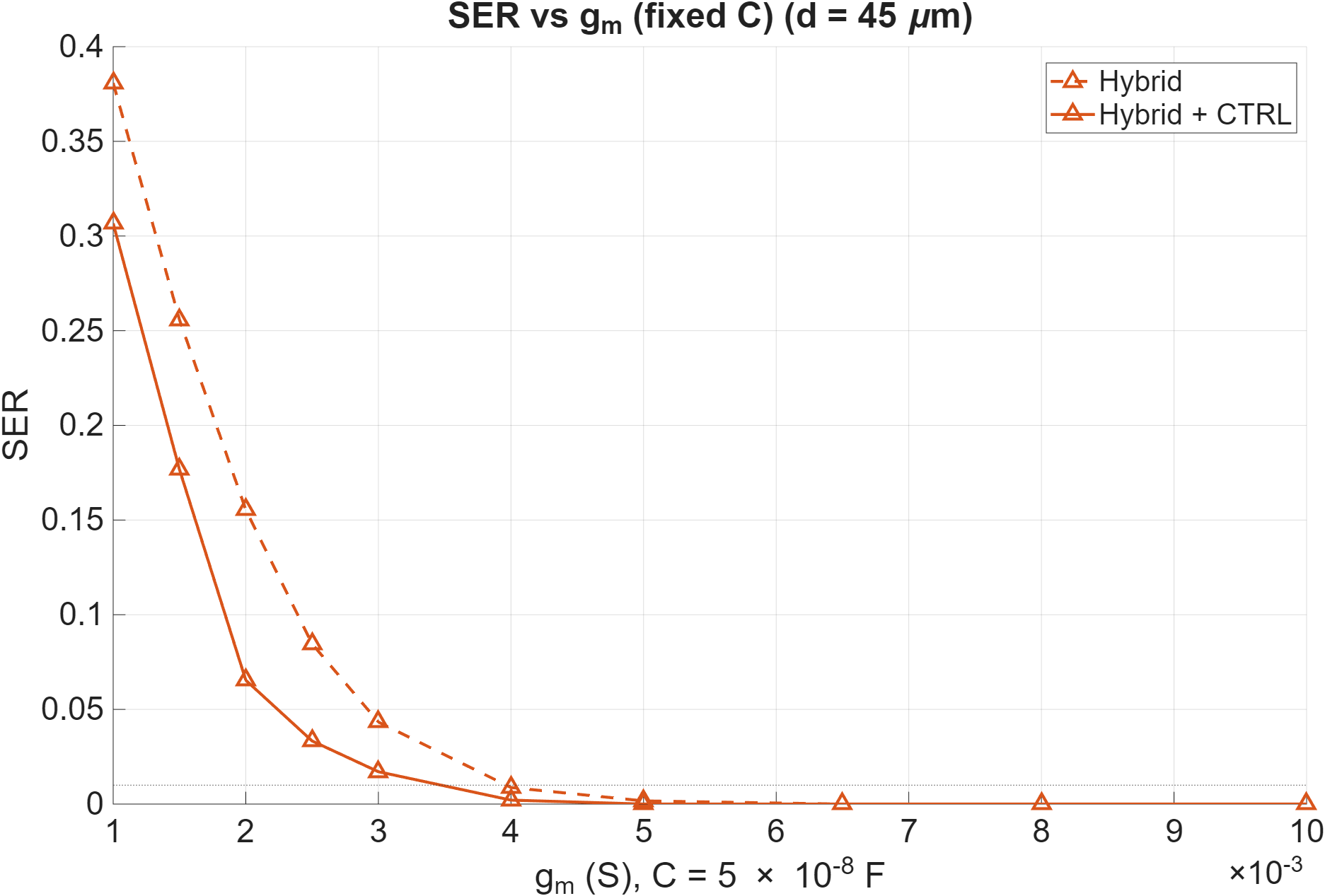}
    \label{fig:ser_vs_gm}
  }
  \hfill
  \subfloat[SER versus $C_{\mathrm{tot}}$ at fixed $g_m=\SI{5}{mS}$.]{
    \includegraphics[width=0.48\textwidth]{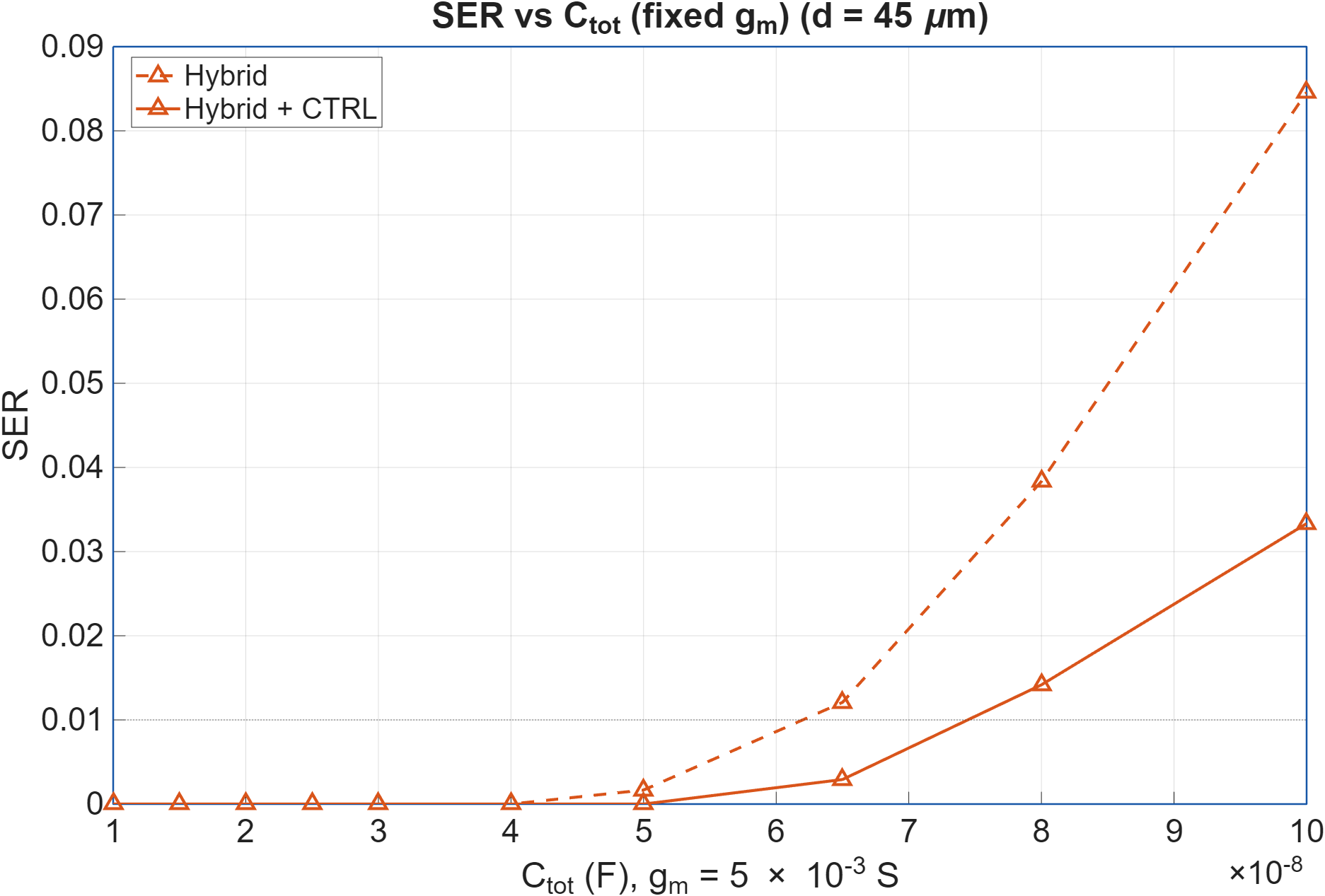}
    \label{fig:ser_vs_C}
  }\\[1mm]
  \subfloat[2-D SER regions with CTRL off.]{
    \includegraphics[width=0.48\textwidth]{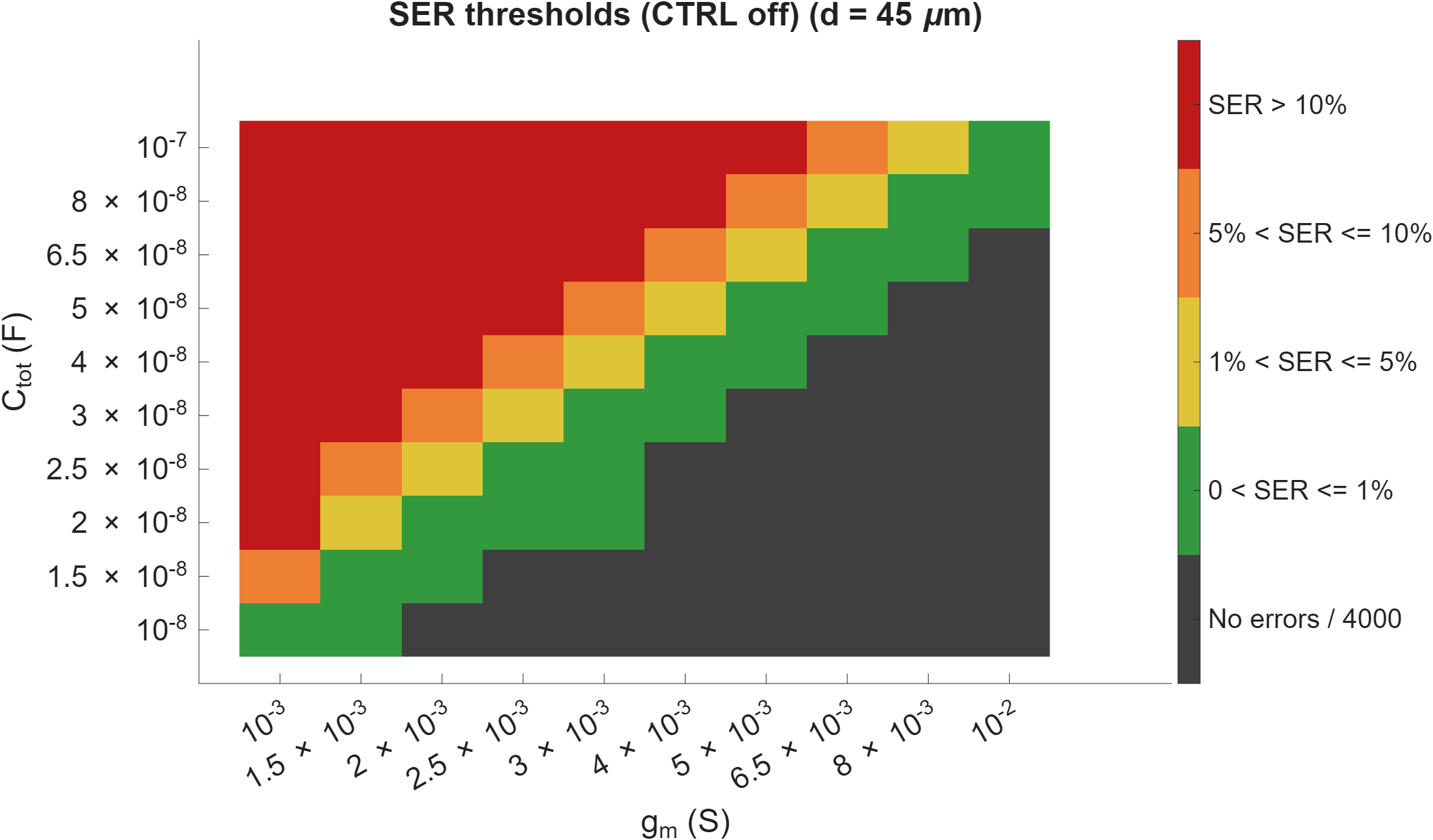}
    \label{fig:heatmap_ctrl_off}
  }
  \hfill
  \subfloat[2-D SER regions with CTRL on.]{
    \includegraphics[width=0.48\textwidth]{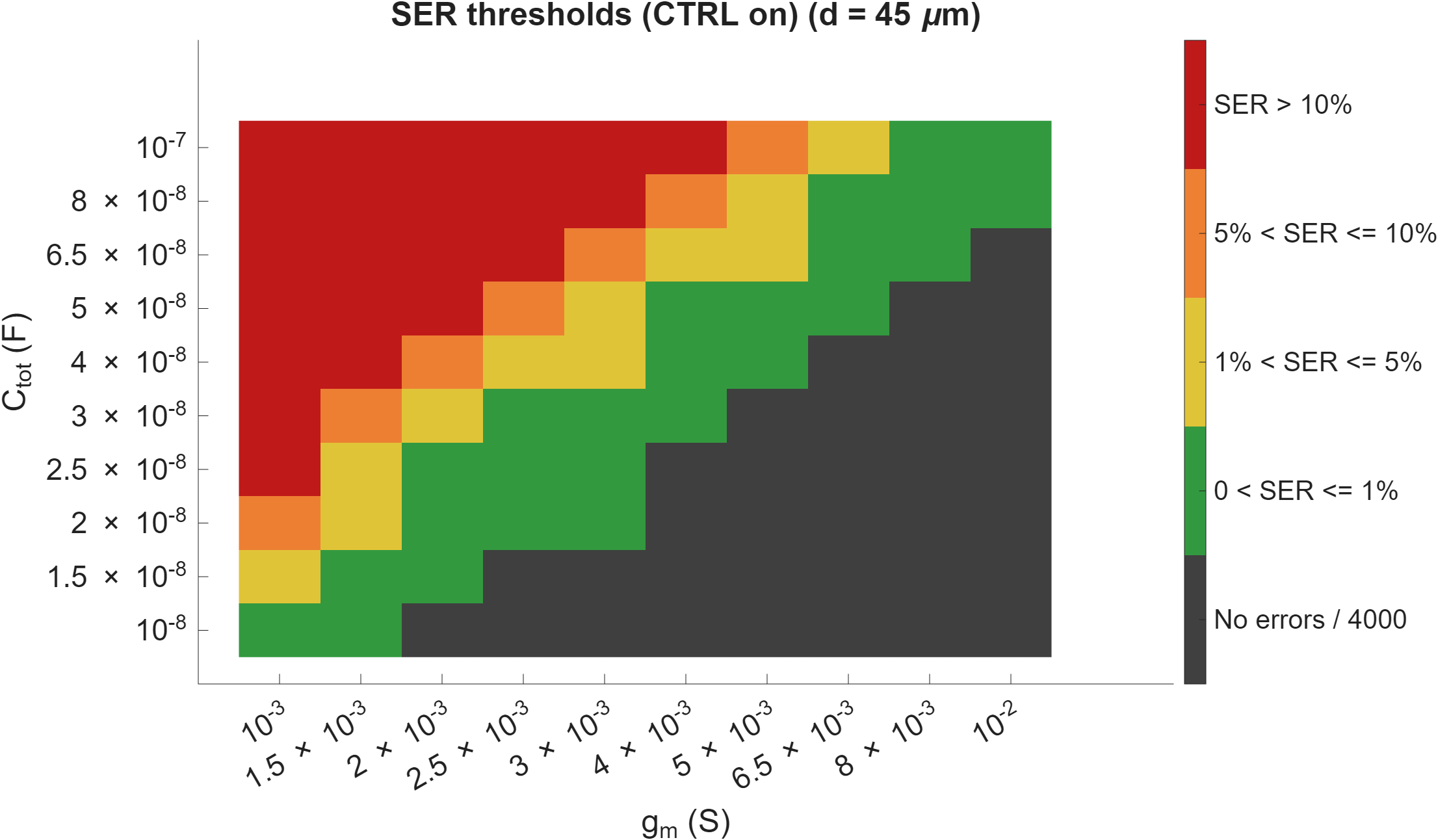}
    \label{fig:heatmap_ctrl_on}
  }
  \caption{Device-level operating envelope for Hybrid reception at $r=\SI{45}{\micro\meter}$. The heatmaps report the sampled $(g_m,C_{\mathrm{tot}})$ region in terms of SER thresholds and no-observed-error points under the simulated run budget.}
  \label{fig:device_results}
\end{figure*}

Figure~\ref{fig:device_results} translates the communication results into device-level design rules. In Fig.~\ref{fig:device_results}(a), increasing $g_m$ at fixed $C_{\mathrm{tot}}=\SI{50}{nF}$ monotonically lowers the Hybrid SER, as expected from the larger signal current in \eqref{eq:oect_signal}. Control referencing is most valuable near the boundary between usable and unusable operating points: at $g_m=\SI{2}{mS}$, the SER decreases from 0.155 to 0.065 with CTRL, whereas at $g_m\ge\SI{6.5}{mS}$ no symbol errors are observed within the simulated budget for either setting.

Figure~\ref{fig:device_results}(b) shows the complementary capacitance sweep at fixed $g_m=\SI{5}{mS}$. Larger $C_{\mathrm{tot}}$ reduces the effective gate perturbation per binding event and therefore degrades detection. At $C_{\mathrm{tot}}=\SI{100}{nF}$, CTRL reduces the SER from $8.46\times 10^{-2}$ to $3.33\times 10^{-2}$. At \SI{65}{nF}, it reduces SER from $1.21\times 10^{-2}$ to $2.92\times 10^{-3}$.

The 2-D threshold maps in Fig.~\ref{fig:device_results}(c) and (d) summarize the feasible design envelope at $r=\SI{45}{\micro\meter}$. With CTRL off, 55 of 100 sampled $(g_m,C_{\mathrm{tot}})$ points lie at or below 1\% SER. With CTRL on, this expands to 60 of 100. Over the same grid, the clearly poor region with SER $>10\%$ shrinks from 29 to 22 points. This relaxation is visible along the frontier. At $C_{\mathrm{tot}}=\SI{100}{nF}$, the minimum $g_m$ required to remain below 1\% drops from \SI{10}{mS} without CTRL to \SI{8}{mS} with CTRL. The control channel therefore does not merely improve a single operating point. It widens the useful OECT design space for Hybrid detection.

\subsection{Sensitivity to Correlation, Transport, and Temperature}

\begin{figure*}[!t]
  \centering
  \begin{minipage}[t]{0.31\textwidth}
    \centering
    \subfloat[SER versus pre-subtraction correlation coefficient $\rho$.\label{fig:ser_vs_corr}]{%
      \includegraphics[width=\linewidth]{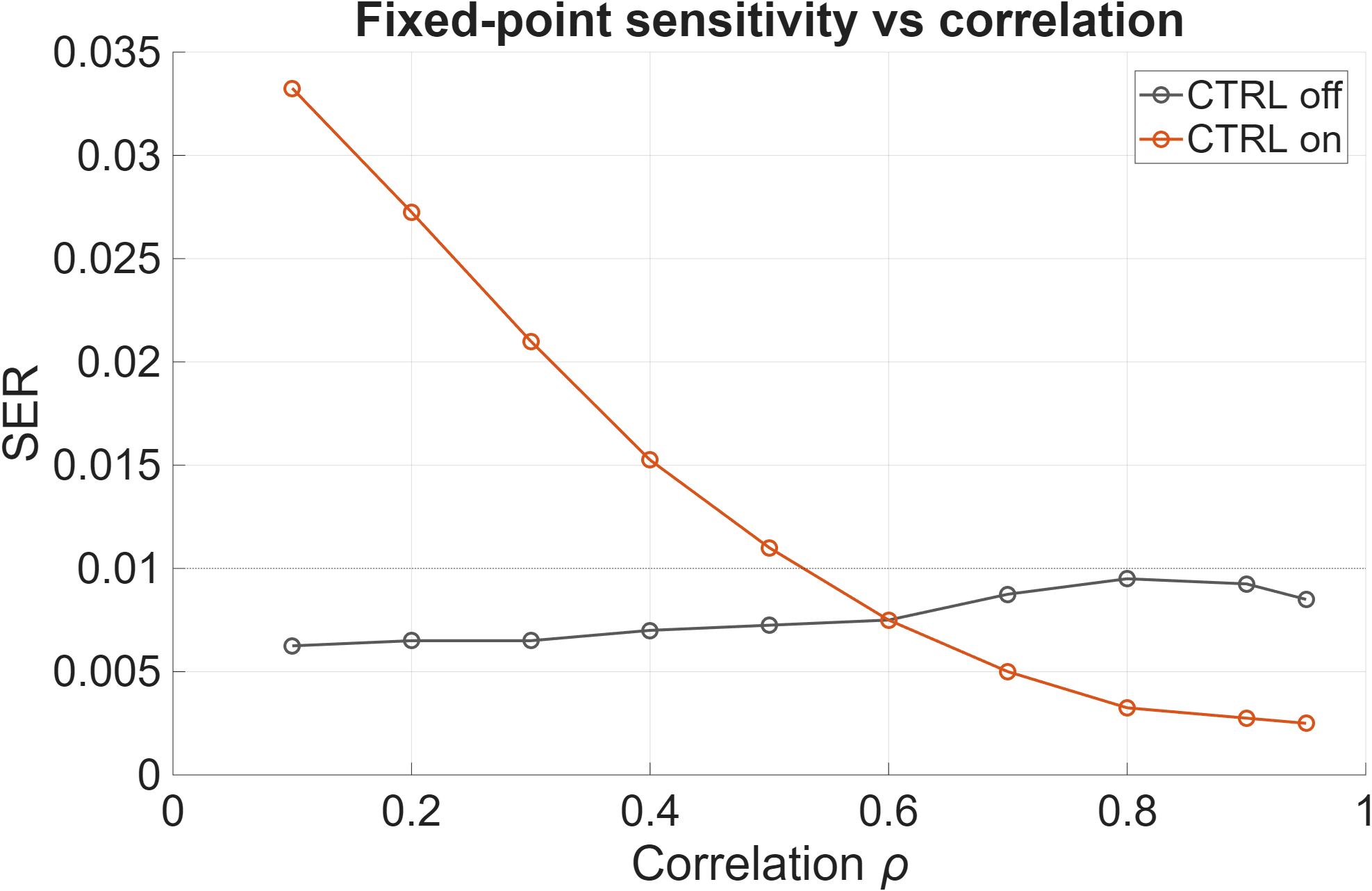}}
  \end{minipage}\hfill
  \begin{minipage}[t]{0.31\textwidth}
    \centering
    \subfloat[SER versus diffusion-scale perturbation.\label{fig:ser_vs_diffusion}]{%
      \includegraphics[width=\linewidth]{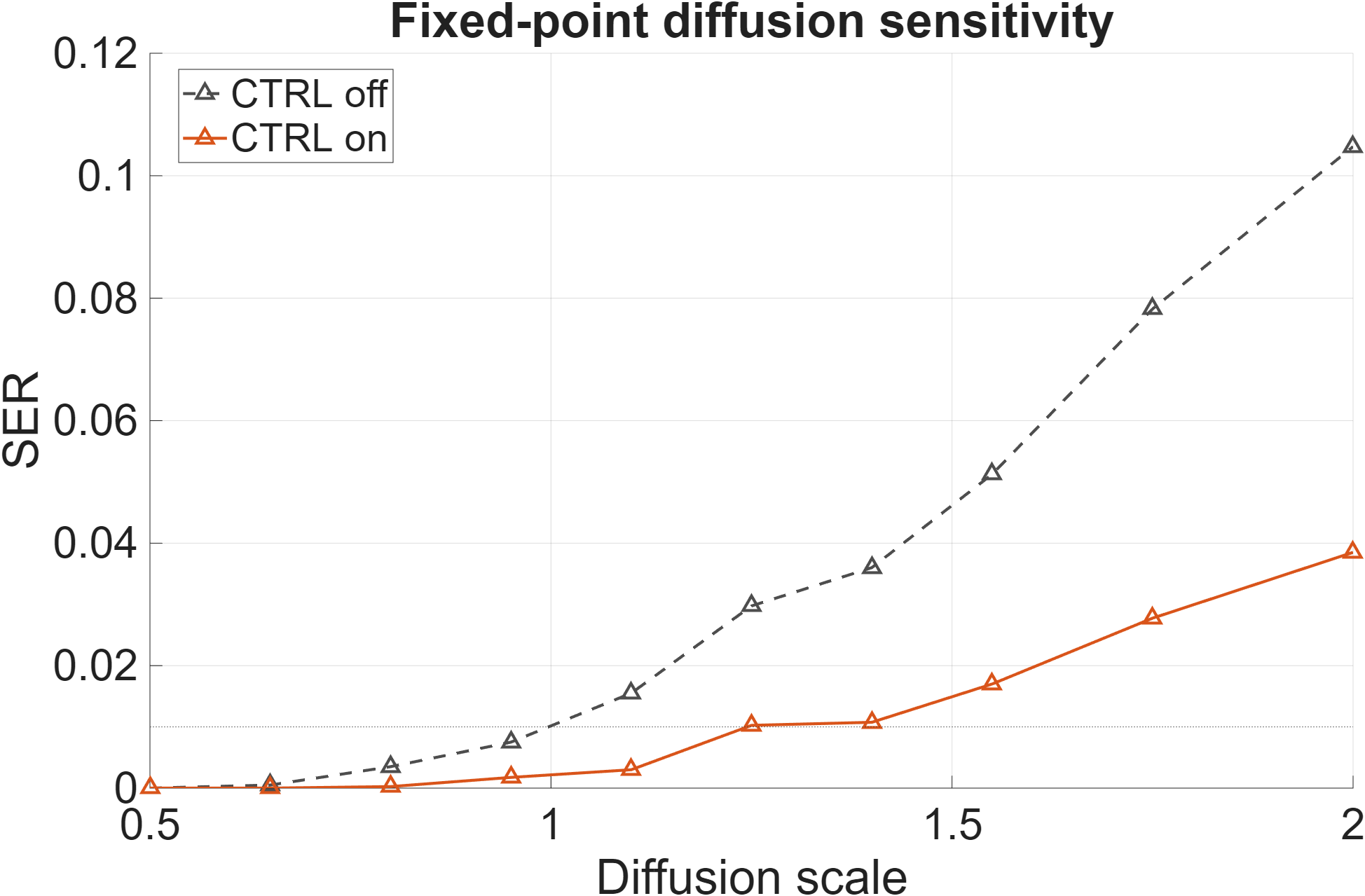}}
  \end{minipage}\hfill
  \begin{minipage}[t]{0.31\textwidth}
    \centering
    \subfloat[SER versus temperature.\label{fig:ser_vs_temp}]{%
      \includegraphics[width=\linewidth]{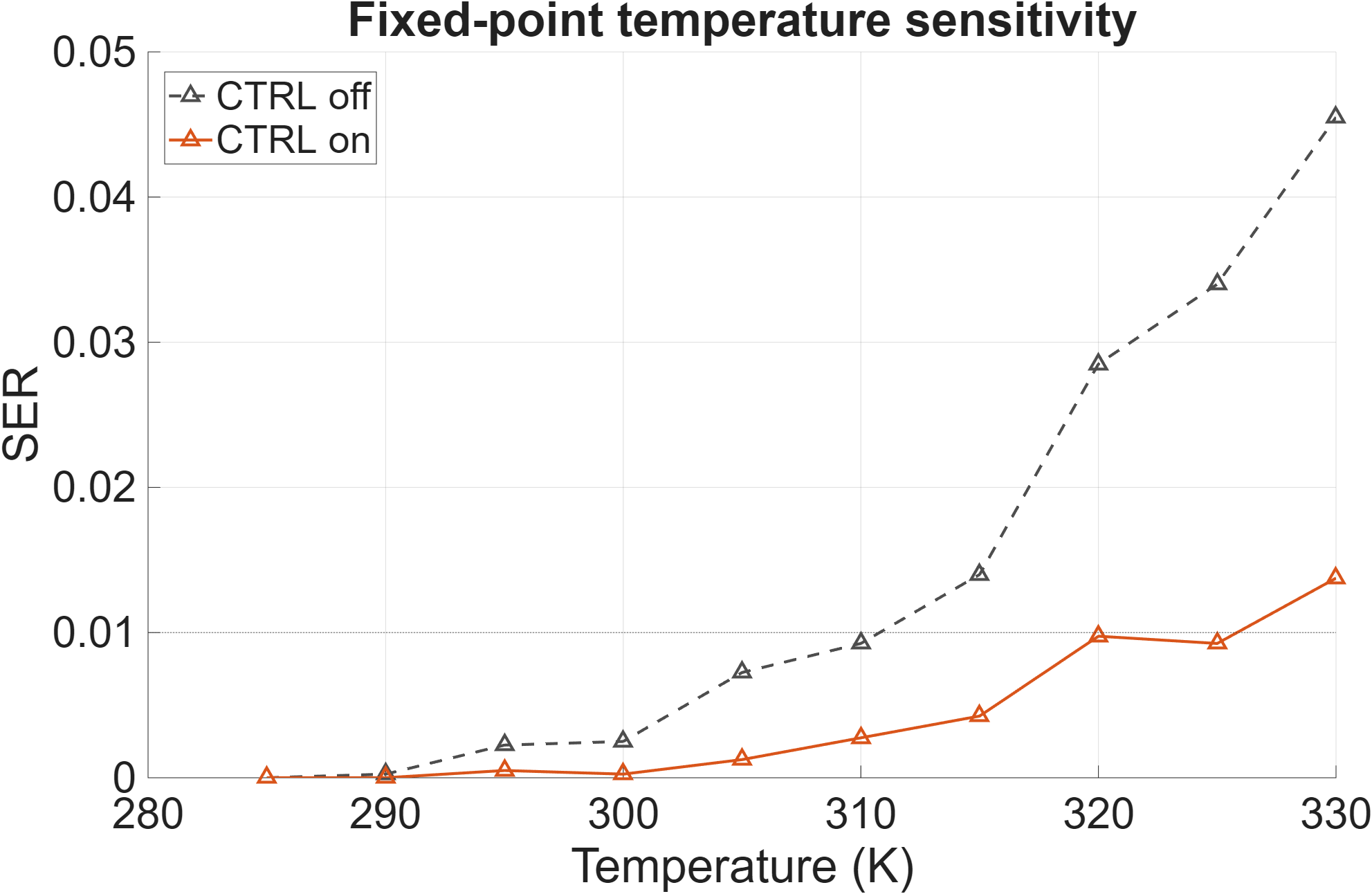}}
  \end{minipage}
  \caption{Robustness of the Hybrid receiver under nuisance-parameter variation. The correlation sweep explains when the control path is beneficial, whereas the diffusion and temperature sweeps quantify sensitivity of the chosen tail-window calibration to transport and environmental changes.}
  \label{fig:robustness_results}
\end{figure*}

Figure~\ref{fig:robustness_results}(a) directly quantifies when control referencing helps. At low pre-subtraction correlation, the reference channel adds more uncorrelated noise than common-mode noise it removes. Accordingly, at $\rho=0.1$ the Hybrid SER rises from $6.25\times 10^{-3}$ without CTRL to $3.33\times 10^{-2}$ with CTRL. The crossover occurs near $\rho\approx 0.6$, where the two curves coincide. This agrees with the approximate condition in \eqref{eq:control_benefit}, which predicts a crossover near $\rho_\star = \tfrac{1}{2}(1+\hat{\sigma}_{q,\mathrm{th}}^2/\hat{\sigma}_{q,\mathrm{lf}}^2)$ for the amplitude branch. Beyond that point, the CTRL path becomes increasingly beneficial, reaching $2.75\times 10^{-3}$ at $\rho=0.9$ versus $9.25\times 10^{-3}$ without CTRL. This crossover corroborates the interpretation from Fig.~\ref{fig:distance_results}: CTRL is valuable in a common-mode-limited regime, not by default.

Figures~\ref{fig:robustness_results}(b) and (c) stress the receiver with transport and environmental perturbations while keeping the detector structure fixed. In the diffusion sweep, increasing the diffusion scale shifts more signal energy toward earlier times and away from the tail-anchored decision window used here. The no-CTRL SER consequently rises from 0 to 0.105 as the diffusion scale increases from 0.5 to 2.0, whereas CTRL limits the same endpoint to 0.0385. In the temperature sweep, the SER grows from 0 at 285 K to 0.0455 at 330 K without CTRL, while control referencing keeps the 330 K value to 0.0138. These are robustness results for the chosen receiver/window calibration rather than universal claims about diffusion or temperature in all MC systems, but they show that the tri-channel architecture retains a clear margin under plausible nuisance variations.

\section{Discussion}
\label{sec:discussion}

\subsection{What the Control Channel Actually Does}

In this paper, control referenced refers specifically to subtracting the simultaneously measured \CTRL current or charge from a selective-channel measurement over the same symbol window before an amplitude decision. The \CTRL channel is hydrogel matched and shares the same electrolyte and gate bias, but it has no aptamer and therefore no intended analyte-dependent binding signal. It is not used as a third communication axis, and it is not part of the MoSK statistic. Its role is to estimate shared baseline motion, drift, flicker, and other correlated low-frequency fluctuations that would otherwise corrupt the amplitude threshold.

The numerical results show that this role is highly specific. Figure~\ref{fig:baseline_results}(b) demonstrates that the Hybrid receiver is not primarily limited by species identification. It is limited by the amplitude branch once a second bit is added. The correlation sweep in Fig.~\ref{fig:robustness_results}(a) then explains why the CTRL path helps only in some regimes: subtraction is advantageous only when the rejected common-mode component is larger than the extra uncorrelated noise introduced by the reference channel. Equation~\eqref{eq:control_benefit} makes that tradeoff explicit and predicts a crossover correlation that approaches 0.5 when low-frequency nuisance dominates. The observed crossover near 0.6 is therefore consistent with a non-negligible but subdominant thermal contribution at the chosen operating point. This is why the benefit is weak or negative at very short range, becomes strong as the signal weakens with distance, and only marginally affects the MoSK component. The third channel is therefore justified as a matched nuisance estimator for the Hybrid amplitude bit rather than as a generic denoising mechanism.

\subsection{Design Implications for Organoid-Facing Receivers}

Several receiver-design rules follow from Figs.~\ref{fig:distance_results}--\ref{fig:robustness_results}. First, the organoid-to-gate gap should be kept as small as the experimental geometry allows, because the transition from near-error-free to control-limited operation occurs rapidly once the gap exceeds roughly 45--\SI{60}{\micro\meter} under the present parameters. Second, if shared low-frequency drift is expected, as is common in electrolyte-gated measurements, the receiver should reserve area for a matched control pixel. The control channel is most useful when the front-end operates close to its SNR limit, which is exactly the regime relevant for minimally invasive, low-bias organoid readout.

Third, device and timing co-design matter as much as chemistry. Higher $g_m$, lower $C_{\mathrm{tot}}$, and an intermediate symbol period under ISI all enlarge the feasible operating envelope. Under the present ISI parameters, the useful Hybrid symbol periods are on the order of $10^2$ s, which corresponds to raw rates on the order of tens of bits per hour. The intended use case is therefore slow chemical-state readout and eventual closed-loop bioelectronic control, not a high-rate molecular data link. Finally, the Hybrid format should be read as a purposeful abstraction toward richer organoid interfacing, not as a literal model of neural coding. Organoid state is not conveyed by a single molecule identity or a single concentration level alone. Future interfaces will likely need joint information in species, amplitude, and timing. In that sense, Hybrid is valuable not because it is the most complex scheme possible, but because it is the simplest scheme in this study that already forces the receiver to resolve more than one chemical feature on the same front-end.

\subsection{Comparison with Prior MC Receiver Literature}

Receiver-centric MC studies have previously established physical design rules for nanoscale biosensor receivers and single-axis BioFET receivers, and more recent work has examined practical graphene receivers, organoid-focused biosensor receivers, frequency-domain detection with cross-reactive receptors, and control-theoretic adaptive thresholding~\cite{kuscu2019survey,kuscu2016physical,kuscu2016sinw,kuscu2021graphene,zhang2025organoid3d,abdali2024freqmodel,civas2024freqdetect,ni2025artrx}. Because geometries, media, targets, and reported metrics differ substantially across these works, Table~\ref{tab:lit_compare} is intended as a qualitative positioning table rather than a direct numerical ranking. Hybrid modulation itself is not new in MC. Prior work has combined concentration with timing~\cite{gursoy2021conctime} and MoSK with CSK in multi-link settings~\cite{sabu2019hybridmoskcsk}. The contribution here is therefore not the introduction of Hybrid modulation as a generic alphabet, but the receiver-architectural question of when a matched control axis changes the feasibility of an identity-plus-amplitude Hybrid symbol on a tri-channel OECT front-end.

\begin{table*}[!t]
  \centering
  \caption{Representative receiver-centric MC studies and closest antecedents in relation to the present work.}
  \label{tab:lit_compare}
  \scriptsize
  \setlength{\tabcolsep}{3pt}
  \renewcommand{\arraystretch}{1.00}
  \begin{tabularx}{\textwidth}{@{}p{0.11\textwidth}p{0.16\textwidth}p{0.15\textwidth}p{0.20\textwidth}p{0.10\textwidth}X@{}}
    \toprule
    Ref. & Receiver platform & Axes / reference & Main focus & Evidence & Relation to this work \\
    \midrule
    \cite{kuscu2016physical} & Nanoscale biosensor receiver abstraction & Single axis, no matched reference & Physical design rules, sensitivity, and noise tradeoffs & Analysis & Foundational receiver design study, but not OECT based and not multi-axis. \\
    \cite{kuscu2016sinw} & SiNW BioFET receiver & Single axis, no matched reference & Unified MC + BioFET SNR/SEP model & Analysis & Single-axis electrical receiver, no tri-channel front-end or Hybrid detector. \\
    \cite{kuscu2021graphene} & Graphene BioFET microfluidic receiver & Single DNA axis & Experimental receiver testbed and difference-based detection under ISI & Experiment & Practical MC receiver, but not matched-reference multi-species sensing and not organoid-like restricted diffusion. \\
    \cite{zhang2025organoid3d} & 3D biosensor receiver for organoids & Single axis, no matched control & Organoid-focused receiver modeling & Analysis + simulation & Closest organoid antecedent. The present work instead studies a tri-channel OECT front-end with a hydrogel-matched control axis and Hybrid identity/amplitude decoding. \\
    \cite{abdali2024freqmodel} & Graphene BioFET microfluidic model & Single axis, no matched reference & Frequency-domain channel/receiver model & Analysis + simulation & Microfluidic end-to-end modeling, no matched control axis and no OECT front-end. \\
    \cite{civas2024freqdetect} & BioFET with cross-reactive receptors & Multiple ligands, no matched reference & Frequency-domain interference rejection & Analysis + simulation & Addresses biochemical cross-talk rather than common-mode electrical drift. \\
    This work & Tri-channel OECT (\DA, \HT, \CTRL) & Two selective axes + matched control & Hybrid identity/amplitude detection, LoD, timing, and device envelope & Monte Carlo & Organizes MoSK, CSK-4, and Hybrid around a common OECT front-end in an organoid-like setting. \\
    \bottomrule
  \end{tabularx}
\end{table*}

Complementing the organoid-focused receiver analysis in \cite{zhang2025organoid3d}, the present paper addresses a different architectural question. Zhang and Akan study a 3D biosensor receiver for Internet-of-Organoids settings, whereas here we analyze a tri-channel OECT front-end with two selective axes and a hydrogel-matched control pixel. Compared with the single-axis analyses in \cite{kuscu2016physical,kuscu2016sinw,abdali2024freqmodel}, the contribution here is not merely a change of device material from FET to OECT. The central addition is a tri-channel matched-reference front-end together with a detector structure that separates molecule identity from control-referenced amplitude on the same receiver. Compared with ART-Rx~\cite{ni2025artrx}, which focuses on symbol-by-symbol threshold adaptation in a diffusion-based BioFET-oriented MC receiver, our emphasis is shifted from adaptive decision control to front-end architecture and matched control referencing for multi-species Hybrid decoding. Compared with the experimental graphene receiver in \cite{kuscu2021graphene}, the present study targets organoid-like restricted extracellular transport and explicitly quantifies how a matched control axis changes the error budget, LoD, timing choice, and device envelope. Compared with frequency-domain interference mitigation using cross-reactive receptors \cite{civas2024freqdetect}, our emphasis is different: we target common-mode electrical nuisance and multimodal identity-plus-amplitude decoding rather than biochemical cross-talk rejection. To the best of our knowledge, no prior receiver-centric MC study has combined a tri-channel OECT front-end, a hydrogel-matched control pixel, and a Hybrid identity-plus-amplitude detector under organoid-like restricted transport and shared low-frequency nuisance.

\subsection{Limitations and Next Steps}

This study is intentionally receiver centric. It adopts an effective point emitter with equal nominal separation to the three channels and absorbs off-target binding and analyte cross-reactivity into effective coupling and noise parameters rather than a full chemical interference matrix. The control path is treated as hydrogel matched, and long-duration adaptive recalibration or device mismatch is outside the present scope. The main conclusion is therefore architectural: a matched control axis materially changes the feasibility of a 2-bit Hybrid receiver under common-mode nuisance. Absolute SER and LoD values remain tied to the baseline scenario of Table~\ref{tab:sim-params}. Future work can extend the same framework to unequal channel spacing, control mismatch, richer cross-reactivity models, and collaborative experimental validation of tri-channel OECT front-ends.

\raggedbottom
\section{Conclusion}
\label{sec:conclusion}

This paper introduced a control-referenced tri-channel OECT receiver for Hybrid molecular communication with brain organoid interfaces, with DA and 5-HT selective channels plus a hydrogel-matched \CTRL channel. The main result is mechanistic. The Hybrid receiver is limited primarily by amplitude errors, and the matched control path specifically suppresses that branch under common-mode low-frequency nuisance. At the baseline \SI{45}{\micro\meter} spacing, the calibrated no-ISI front-end benchmark reaches the 1\% SER region at $1.55\times 10^4$ molecules/symbol and achieves an LoD of 11866 molecules/symbol, lower than CSK-4+CTRL while retaining a 2-bit symbol. The benefit grows over much of the medium-to-long-distance regime, and additional sweeps identify an intermediate symbol period under ISI together with a wider $(g_m,C_{\mathrm{tot}})$ operating envelope when CTRL is used. Under the present ISI parameters, that operating region corresponds to raw rates on the order of tens of bits per hour, consistent with slow chemical-state readout and eventual closed-loop bioelectronic control rather than a high-rate molecular data link. Taken together, the results support the tri-channel OECT receiver as an architecture-level receiver-design step toward multimodal, low-bias chemical readout for brain organoids and motivate future collaborative validation together with richer identity-amplitude-timing signaling.


\bibliographystyle{IEEEtran}
\bibliography{references}

\begin{IEEEbiography}[{\includegraphics[width=1in,height=1.25in,clip,keepaspectratio]{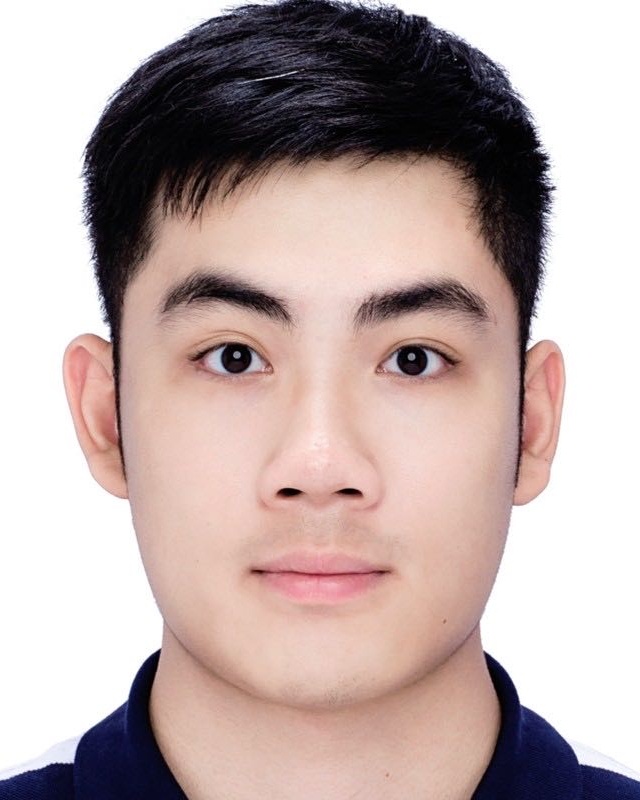}}]
{Hongbin Ni}
(Member, IEEE) received the B.Eng. degree in Mechatronic Engineering from The University of Manchester, U.K., in 2021, and the M.Sc. degree in Biomedical Engineering from Imperial College London, U.K., in 2022. He is currently working toward the Ph.D. degree in Engineering with the Internet of Everything Group, Department of Engineering, University of Cambridge, U.K. His research interests include molecular communications and its applications to neural interfaces.
\end{IEEEbiography}

\begin{IEEEbiography}[{\includegraphics[width=1in,height=1.25in,clip,keepaspectratio]{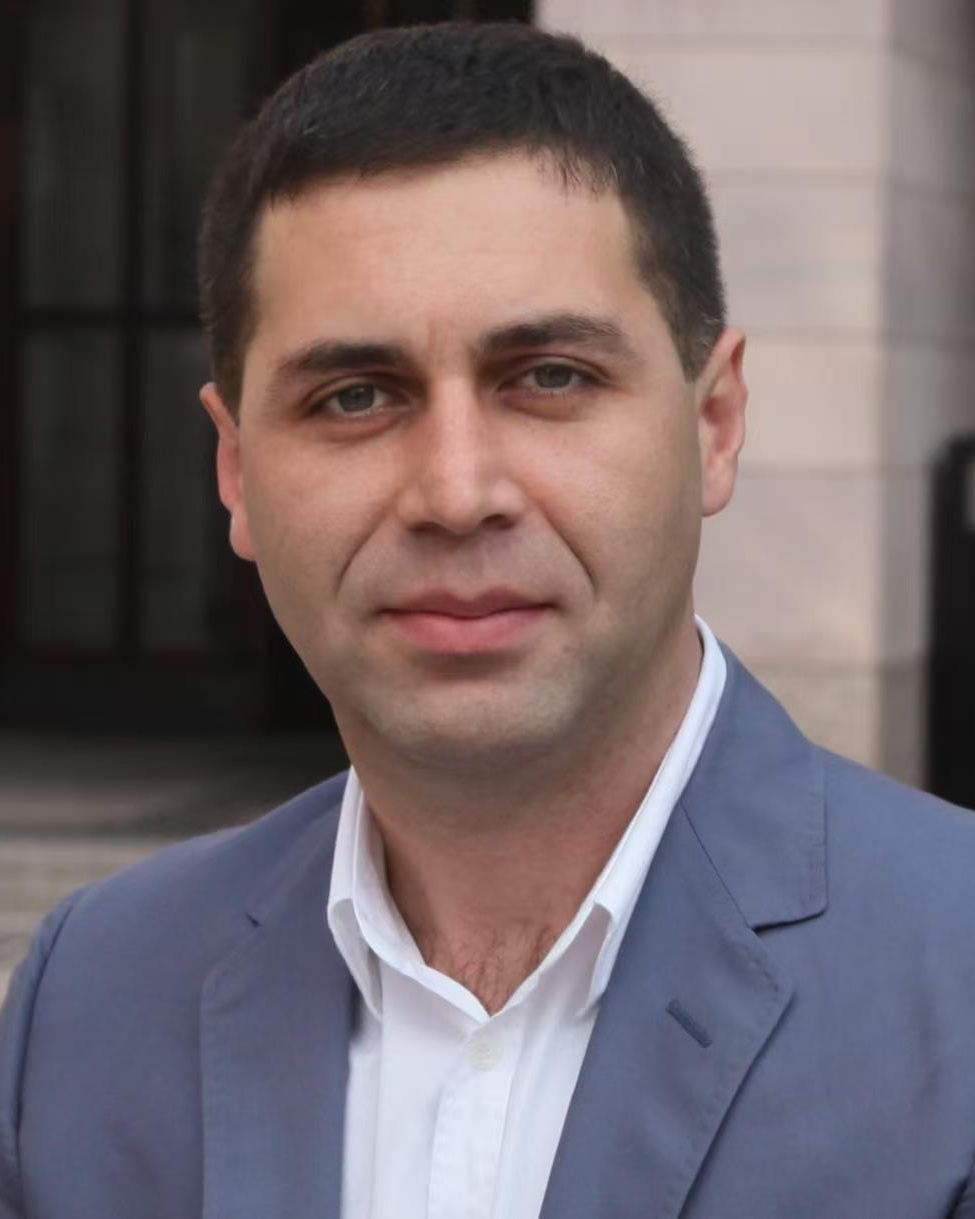}}]
{Ozgur B.~Akan}
(Fellow, IEEE) received the Ph.D. degree from the School of Electrical and Computer Engineering, Georgia Institute of Technology, Atlanta, in 2004. He is currently the Head of the Internet of Everything Group, Department of Engineering, University of Cambridge, U.K., and the Director of the Centre for Next-Generation Communications, Koç University, Türkiye. His research interests include wireless, nano, and molecular communications and Internet of Everything.
\end{IEEEbiography}

\end{document}